\documentclass[12pt,a4paper]{article}

\usepackage{amsmath,amssymb,slashed,cancel}
\usepackage{cite,tabularx}
\usepackage{subcaption}
\usepackage{graphicx,xcolor}
\usepackage[normalem]{ulem}
\usepackage{mathtools}
\usepackage{enumerate}
\usepackage{multirow}
\usepackage{bm}
\usepackage{braket}
\usepackage{comment}

\allowdisplaybreaks

\usepackage[height=9.5in,width=6.4in]{geometry}

\newcommand\package[2][\relax]{\texttt{#2\ifx#1\relax\relax\relax\else\,\linebreak[0]#1\fi}}

\newcommand\dd{\mathrm{d}}

\numberwithin{equation}{section} 
\def\beq#1\eeq{\begin{align}#1\end{align}}

\definecolor{BlueViolet}{rgb}{0.2, 0.00, 0.7}
\definecolor{Blue}{rgb}{0.15, 0.00, 0.9}
\usepackage[colorlinks=true,linkcolor=Blue,citecolor=Blue,urlcolor=BlueViolet]{hyperref}

\begin{document}
\begin{titlepage}
\setcounter{page}{0} 

\begin{center}

\hfill {\tt YITP-25-187}\\

\vskip .45in

\begingroup
\centering
\large\bf
Cosmological Bounds on Scotogenic Model\\
with Asymmetric Mediator
\endgroup

\vskip .4in

\renewcommand{\thefootnote}{\fnsymbol{footnote}}
{
Kento Asai$^{(a,b)}$\footnote{
  \href{mailto:kento.asai@yukawa.kyoto-u.ac.jp}
  {\tt kento.asai@yukawa.kyoto-u.ac.jp}},
Seishi Enomoto$^{(c,d)}$\footnote{
  \href{mailto:enomoto@ip.kyusan-u.ac.jp}
  {\tt enomoto@ip.kyusan-u.ac.jp}},
Takuya Hirose$^{(c)}$\footnote{
  \href{mailto:t.hirose@ip.kyusan-u.ac.jp}
  {\tt t.hirose@ip.kyusan-u.ac.jp}}, and 
Masato Yamanaka$^{(e)}$\footnote{
  \href{mailto:masato.yamanaka.48@hosei.ac.jp}
  {\tt m.yamanaka.km@cc.it-hiroshima.ac.jp }}
}

\vskip 0.4in

\begingroup\small
\begin{minipage}[t]{0.9\textwidth}
\centering\renewcommand{\arraystretch}{0.9}
{\it
\begin{tabular}{c@{\,}l}
$^{(a)}$
& Yukawa Institute for Theoretical Physics, Kyoto University, \\ & Kyoto 606--8502, Japan \\[2mm]
$^{(b)}$
& Institute for Cosmic Ray Research (ICRR), The University of Tokyo, \\
& Kashiwa, Chiba 277--8582, Japan \\[2mm]
$^{(c)}$
& Faculty of Science and Engineering, Kyushu Sangyo University,  \\
& Fukuoka 813--8503, Japan \\[2mm]
$^{(d)}$
& Department of Physics, Faculty of Engineering Science, \\
& Yokohama National University, Yokohama 240--8501, Japan \\[2mm]
$^{(e)}$
& Department of Global Environment Studies, \\
& Hiroshima Institute of Technology, Hiroshima, 731--5193, Japan \\[2mm]

\end{tabular}
}
\end{minipage}
\endgroup

\end{center}

\vskip .2in

\begin{abstract}
\noindent
We study cosmological constraints on the asymmetric mediator scenario, a variant of the scotogenic model that addresses the origins of neutrino masses, dark matter (DM), and the baryon asymmetry.
An SU(2)$_L$ doublet scalar $\eta$ mediates between the visible and dark sectors, while a singlet scalar $\sigma$ serves as the DM candidate.
We evaluate the DM relic abundance by solving the Boltzmann equations including $\eta$ decay and scattering processes prior to the freeze-out of the $\eta$ asymmetry, and show Big Bang nucleosynthesis constraints from late-time $\eta$ decays.
Combining the DM abundance and BBN bounds, we find the favored parameter space of this model, for 
instance, the mediator masses of $m_\eta \lesssim \mathcal{O}(10)$\,TeV.

\end{abstract}
\end{titlepage}
\setcounter{page}{1}
\renewcommand{\thefootnote}{\#\arabic{footnote}}
\setcounter{footnote}{0}


\section{Introduction}
\label{sec:introduction}

Some fundamental issues remain as long-standing puzzles in particle physics, e.g., the nonzero but extremely tiny neutrino masses, the baryon asymmetry of the universe 
(BAU)~\cite{Planck:2018vyg}, the nature of dark matter (DM), and so on. 
Another puzzle that warrants attention is the present-day energy densities of DM ($\rho_{\text{DM}}$) 
and baryonic matter ($\rho_{\text{B}}$). Cosmological and astrophysical observations have shown that 
the DM energy density today is relatively close to the baryonic one, 
$\rho_{\text{DM}} \simeq 5.4 \rho_{\text{B}}$~\cite{WMAP:2012nax,Planck:2018vyg}. 
Insight from the theory of everything leads us to envision a unified description of these 
issues within the context of a particle physics model beyond the Standard Model (SM). 

A well-motivated unified picture to solve the above puzzles is the seesaw mechanism~\cite{Minkowski:1977sc, 
Yanagida:1979as, Gell-Mann:1979vob, Mohapatra:1979ia} induced by additional SM singlet matter fields with heavy Majorana masses $M_i$ and the Dirac masses $m_{\nu_\alpha}^D$ of the neutrinos.
The large mass hierarchy between the Majorana and the Dirac mass of the neutrinos realizes the tiny neutrino masses as $m_{\nu} = - m_{\nu}^D M^{-1} (m_{\nu}^D)^T$.
Moreover, the seesaw mechanism offers a big bonus to cosmology, i.e., leptogenesis~\cite{Fukugita:1986hr}.
The CP-violating decays of the singlet matter fields in the early universe give rise to the lepton asymmetry, which is converted into the baryon asymmetry via the electroweak sphaleron process~\cite{Kuzmin:1985mm}.

An alternative framework for such a unified description is the scotogenic model~\cite{Tao:1996vb,Ma:2006km}, which is a subclass of seesaw scenarios.
The scotogenic model extends the SM by introducing heavy singlet neutrinos $N_i$ and an SU(2)$_L$ doublet scalar field $\eta=(\eta^+,\eta^0)^T$, which  radiatively generates the observed masses and mixings of the active neutrinos.
When the neutral component of $\eta$ is lighter than the charged one and the heavy singlet neutrinos, it is stabilized by a $Z_2$ parity and can be a candidate for the DM.
Furthermore, the generation of matter-antimatter asymmetry has recently been studied in this context~\cite{Hugle:2018qbw,Borah:2018uci,Bose:2024bnp,Racker:2024fpn}. 
Combining these results with other phenomenological observables, the characteristics of the model can be specified with high precision.

Let us turn to the issue of the closeness between the energy densities of DM and baryonic matter.
A well-known scenario to account for their closeness is the 
asymmetric DM (ADM)~\cite{Nussinov:1985xr,Barr:1990ca,Barr:1991qn,Dodelson:1991iv, Kaplan:1991ah,Kuzmin:1996he,Foot:2003jt,Foot:2004pq,Hooper:2004dc,Kitano:2004sv,
Gudnason:2006ug,Kaplan:2009ag,Davoudiasl:2012uw,Petraki:2013wwa,Zurek:2013wia}, based on the simple assumption that the BAU originates from an asymmetry between the DM particle and its antiparticle. 
There are two types of DM asymmetry generation mechanisms.
One type is the sharing mechanism: an asymmetry is generated at the beginning, then the DM and SM sectors share it through exotic interactions.
Another type is the cogenesis mechanism: the asymmetries of baryonic matter and DM are generated simultaneously.

In recent years, a well-motivated UV scenario for the ADM has been proposed~\cite{Asai:2022vat}, based on a model that combines the scotogenic framework with the cogenesis mechanism.
The particle content is extended from that of the SM by introducing $Z_2$-odd three Majorana fields $N_i$ and a $Z_2$-odd SU(2)$_L$ doublet scalar field $\eta$, as is the case with conventional scotogenic models, and a SM singlet scalar $\sigma$.
The additional contents $N_i$ and $\eta$ radiatively generate the observed masses and mixing angles in the active neutrino sector. 
The another content $\sigma$ plays the role of DM. 
The BAU is accounted for by the leptogenesis mechanism. However, the Majorana fields decay into SU(2)$_L$ doublet leptons with $\eta$ bosons rather than the SM Higgs bosons. 
Thus, except for the sign, the $\eta$-$\eta^*$ asymmetry is equal to the lepton-antilepton asymmetry generated by the CP-violating decay of the Majorana fields.
This asymmetry is stored in the dark sector due to the conservation of the $Z_2$-parity as long as a CP-violating process between $\eta$ and the SM Higgs boson is suppressed. 
After the $\eta$ asymmetry generation, the $\eta$ and $\eta^*$ particles annihilate each other until the symmetric component is depleted. As a result, the relic number density of $\eta$ or $\eta^*$ is equal to the lepton asymmetry. After this pair-annihilation, $\eta$ or $\eta^*$ decays into the singlet scalar $\sigma$ and the SM particles.
Consequently, the DM number density is automatically equal to the lepton asymmetry and thus relates to the BAU. We refer to this as the asymmetric mediator scenario.

In this work, we investigate the cosmological consequences of the $\eta$ field. 
The charged component $\eta^\pm$ of the doublet $\eta$ can be long-lived due to the mass degeneracy with 
the neutral component $\eta^0$, 
$m_{\eta^\pm}\sim m_{\eta^0}$,
as well as the tiny coupling to the DM field ($\mu$) in the decay. 
In general, long-lived exotic particles can play important roles in the Big Bang Nucleosynthesis (BBN). 
In the case that such particles decay and/or annihilate before or during BBN, their energetic decay or 
annihilation products can modify the neutron-to-proton ratio or destroy primordial light elements. 
Bounds have been placed on their lifetimes and decay or annihilation channels to avoid spoiling the 
successful predictions of standard BBN~\cite{Reno:1987qw, Kawasaki:2004qu, Kawasaki:2020qxm, Kohri:2001jx, 
Kawasaki:2017bqm, Jedamzik:2006xz, Khlopov:1984pf, Lindley:1984bg, Forestell:2018txr, Dimopoulos:1988ue}. 
Such non-standard nuclear reactions have also been studied as possible solutions to the so-called lithium 
problems~\cite{Jedamzik:2004er, Cumberbatch:2007me, Koren:2022axd, Goudelis:2015wpa, Yamazaki:2014fja}, 
which refer to the long-standing discrepancies between the observed and theoretically predicted primordial 
abundances of $^7$Li and $^6$Li~\cite{ParticleDataGroup:2024cfk}.
In addition to decays and scatterings, long-lived negatively charged particles can form bound states with nuclei, triggering catalytic nuclear reactions~\cite{Pospelov:2006sc}.
Non-standard nuclear chains induced by such exotic bound states are extensively studied~\cite{Pospelov:2008ta, Khlopov:2007ic, 
Jittoh:2011ni, Kamimura:2008fx} and shown to open toward successful BBN predictions while 
alleviating the lithium problems~\cite{Kaplinghat:2006qr, Kohri:2006cn, Pospelov:2006sc, Hamaguchi:2007mp, 
Jittoh:2007fr, Jittoh:2010wh, Jittoh:2008eq, Bird:2007ge, Jedamzik:2007cp, Kusakabe:2007fu, Bailly:2008yy}. 
The impact of these exotic processes on BBN strongly depends on the properties of the relic particles, 
such as their lifetimes, decay products, and the mechanism responsible for their longevity. With detailed 
analyses of nuclear reaction networks incorporating updated experimental data~\cite{Boyd:2010kj}, 
the properties of long-lived particles and the underlying particle physics models can be quantitatively 
predicted and constrained.

Another important cosmological aspect related to $\eta$, in addition to BBN, is the relic abundance of DM. 
In this scenario, the DM is produced from $\eta$ decay and can be categorized into two types depending on 
the epoch of production: 
(1) long after the freeze-out of $\eta$ pair annihilation, and
(2) before the freeze-out. 
The previous study~\cite{Asai:2022vat} has considered only the first case, in which the cogenesis mechanism is automatically maintained as mentioned above.
However, the additional contribution in the second case must be carefully examined, as it may disrupt the cogenesis mechanism in this scenario.
The conditions for successful BBN and DM relic abundance could place nontrivial constraints on the mass degeneracy of the charged and neutral $\eta$s, $\delta m_\eta$, and the scalar coupling, $\mu$.

This paper is organized as follows. 
In Sec.~\ref{sec:model}, we review the asymmetric mediator scenario~\cite{Asai:2022vat}. 
We also demonstrate the parameter dependence of $\eta$ lifetime.
In Sec.~\ref{sec:pheno}, we calculate the decay rate of the $\eta$ particle, and the 
cross section relevant for the network of Boltzmann equations of $\eta$. 
In Sec.~\ref{sec:result}, we show the parameter space favored by the successful BBN and relic abundance of DM.
Our concluding remarks are given in Sec.~\ref{sec:summary}.

\section{Scotogenic Model with Singlet Scalar Dark Matter}
\label{sec:model}

In this section, we review the asymmetric mediator scenario~\cite{Asai:2022vat} based on the scotogenic model.
The original scotogenic model comprises the SM content, three generations of the right-handed neutrinos $N_i~(i=1,2,3)$, and an inert scalar doublet $\eta$.
The $Z_2$ parity is assigned even for the SM fields, while odd for $N_i$ and $\eta$.
In addition, we introduce a real singlet scalar $\sigma$ with the $Z_2$ odd parity.
The singlet scalar $\sigma$ is the lightest particle among the $Z_2$-odd particles and plays the role of DM in our model. 
See Table~\ref{tab:Field_content} for the particle content and the corresponding charges. 

\begin{table}[t]
    \centering
    \caption{Particle content and corresponding charges of the fields beyond the SM in the extended scotogenic model.}
    \label{tab:Field_content}
    \begin{tabular}{ll||c|c|c} \hline
        \multicolumn{2}{c||}{Field} & SU(2)$_L$ & U(1)$_Y$ & $Z_2$ \\ \hline
        $N_i \ (i=1,2,3)$ & (Majorana fermions) &  $\mathbf{1}$ & 0 & odd \\ 
        $\eta=\left( \begin{array}{c} \eta^+ \\ \eta^0 \end{array} \right)$ & (Complex scalar) & $\mathbf{2}$ & 1/2 & odd\\
        $\sigma$ & (Real scalar)& $\mathbf{1}$ & 0 & odd \\ \hline
    \end{tabular}
\end{table}

\subsection{Lagrangian}
We denote the SM left-handed lepton doublet and the Higgs scalar doublet as $\ell$ and $H$, respectively.
The Lagrangian of the asymmetric mediator model is given by
\begin{align}
\label{eq:lagrangian}
    \mathcal{L}&=\mathcal{L}_{\text{SM}}+\mathcal{L}_{N}-V(H,\eta,\sigma)~, \\
\label{eq:lagrangian-nu}
    \mathcal{L}_N&=-h_{\alpha i}\bar{\ell}_{\alpha}\tilde{\eta}N_i+\frac{1}{2}M_{i}\bar{N}_{i}N^{c}_{i}+\mathrm{H.c.}~, \\
\label{eq:potential}
    V(H,\eta,\sigma)&= \mu_H^2|H|^2+m_\eta^2|\eta|^2+\frac{1}{2}\mu_\sigma^2\sigma^2 +\frac{1}{2}\lambda_1|H|^4+\frac{1}{2}\lambda_2|\eta|^4+\frac{1}{2}\lambda_3\sigma^4 \nonumber \\
     &\quad +\lambda_4|H|^2|\eta|^2+\lambda_5|H^\dagger\eta|^2+\lambda_6|H|^2\sigma^2+\lambda_7|\eta|^2\sigma^2 \nonumber \\
     &\quad + \frac{1}{2}\lambda_8\left[(H^\dagger\eta)^2+({\rm H.c.})\right]+\frac{1}{\sqrt{2}}\mu\sigma\left[(H^\dagger\eta)+({\rm H.c.})\right]~,
\end{align}
where we denote $\tilde{\eta}\equiv i\sigma_2\eta^*$ with the Pauli matrix $\sigma_2$, and the index $\alpha\,(i)$ labels the flavor (mass) eigenstates.
We also denote the parameters as follows: $h_{\alpha i}$ for a Yukawa coupling responsible for the Dirac mass term of the neutrinos; $M_i$ for the mass eigenvalues of the right-handed neutrinos; $\mu_H^2<0$, $\mu_\eta^2>0$, $\mu_\sigma^2>0$ for mass parameters of each scalar; $\lambda_i \ (i=1,2,\cdots, 8)$ for the dimensionless coupling constants; and $\mu$ for a coupling constant with a mass dimension.
Note that in our setup the Higgs scalar doublet $H$ has a non-zero vacuum expectation value (VEV), while the inert scalar doublet $\eta$ and the singlet scalar $\sigma$ do not.
In the unitary gauge, the components of $\eta$ are represented as
\begin{align}
    H=\left(0,(v+h)/\sqrt{2}\right)^T, \qquad \eta=\left(\eta^+, (\eta_R+i\eta_I)/\sqrt{2}\right)^T, \qquad \eta^-=(\eta^+)^*,
    \label{doublets}
\end{align}
where $v=246$\,GeV is a VEV of the SM Higgs field.
Concerning the couplings in the scalar potential, the coupling constants $\lambda_i$ can be real without loss of generality.
From the perspective of vacuum stability and perturbative unitarity, several couplings satisfy $\lambda_{1,2}>0, \lambda_4+\lambda_5-\lambda_8>-2\sqrt{\lambda_1 \lambda_2}$, and $\lambda_i<4\pi$ for $i=1,2,4,5,8$~\cite{LopezHonorez:2006gr}.
Additionally, we assume $\lambda_6, \lambda_7\ll1$ to avoid constraints from direct detection experiments and thermalization of the DM in the early universe.
Note that if $\lambda_8$ is too small, the observed masses of the active neutrinos cannot be explained, since they are generated radiatively at the one-loop level and are approximately proportional to $\lambda_8$, under the reasonable assumptions $m_\eta \ll M_i$ and $\lambda_8 v^2 \ll m^2_\eta$.
The remaining coupling $\lambda_3$, which governs the self-interacting DM (SIDM), is constrained by observations of the galaxy groups as $\sigma_{\rm SIDM} /m_{\rm DM} \lesssim 0.2 \, {\rm cm}^2/$g  with $\sigma_{\rm SIDM}$ being the self-scattering cross section of the DM \cite{Sagunski:2020spe,Andrade:2020lqq,Eckert:2022qia,K:2023huw}, which corresponds to $\lambda_3\lesssim 36 \times (m_\sigma/{\rm GeV})^{3/2}$.

\subsection{Mass spectrum}
In general, the  CP-even neutral component of $\eta$ and $\sigma$ are mixed in the mass eigenstate.
%
%
Using Eq.~\eqref{doublets}, the mass terms of all scalars are represented as
\begin{align}
    \mathcal{L}_{\rm mass}
     &= \frac{1}{2}\left(\mu_H^2+\frac{3}{2}\lambda_1v^2\right)h^2+\frac{1}{2}\left(m_\eta^2+\frac{1}{2}\left(\lambda_4+\lambda_5-\lambda_8\right)v^2\right)\eta_I^2 +\left(m_\eta^2+\frac{1}{2}\lambda_4v^2\right)\eta^-\eta^+ \nonumber \\
     &\quad +\frac{1}{2}\left(\eta_R \ \sigma \right)\left(\begin{array}{cc}m_\eta^2+\frac{1}{2}\left(\lambda_4+\lambda_5+\lambda_8\right)v^2 & \frac{1}{\sqrt{2}}\mu v \\ \frac{1}{\sqrt{2}}\mu v & \mu_\sigma^2 + \lambda_6 v^2 \end{array} \right)\left(\begin{array}{c}\eta_R \\ \sigma \end{array}\right) \nonumber \\
     &= \frac{1}{2}m_h^2h^2+\frac{1}{2}m_{\eta_I}^2\eta_I^2 +m_{\eta^\pm}^2\eta^-\eta^+ +\frac{1}{2}\left(\tilde{\eta}_R \ \tilde{\sigma} \right)\left(\begin{array}{cc}\tilde{m}_{\eta_R}^2 & \\ & \tilde{m}_\sigma^2 \end{array} \right)\left(\begin{array}{c}\tilde{\eta}_R \\ \tilde{\sigma} \end{array}\right)~,
\end{align}
where
\begin{align}
    m_h^2&=\mu_H^2+\frac{3}{2}\lambda_1v^2~, \qquad m_{\eta_I}^2=m_\eta^2+\frac{1}{2}(\lambda_4+\lambda_5-\lambda_8)v^2~, \\ 
    m_{\eta^\pm}^2&=m_\eta^2+\frac{1}{2}\lambda_4v^2~, \qquad m_{\eta_R}^2=m_\eta^2+\frac{1}{2}(\lambda_4+\lambda_5+\lambda_8)v^2~, \\
    \tilde{m}_{\eta_R}^2&=\frac{1}{2}\left[m_{\eta_R}^2+\mu_\sigma^2+\lambda_6v^2+\sqrt{(m_{\eta_R}^2-\mu_\sigma^2-\lambda_6v^2)^2+2\mu^2v^2}\right]~, \\
    \tilde{m}_\sigma^2&=\frac{1}{2}\left[m_{\eta_R}^2+\mu_\sigma^2+\lambda_6v^2-\sqrt{(m_{\eta_R}^2-\mu_\sigma^2-\lambda_6v^2)^2+2\mu^2v^2}\right]~,
\end{align}
and we define
\begin{align}
    \tilde{\eta}_R=\eta_R\cos\theta+\sigma\sin\theta~, \qquad \tilde{\sigma}=\sigma\cos\theta-\eta_R\sin\theta~.
\end{align}
The mixing angle $\theta$ is satisfied by
\begin{align}
\label{eq:angle}
    \tan2\theta=\frac{\sqrt{2}\mu v}{m^2_{\eta_R}-m^2_{\sigma}}=\frac{\sqrt{2}\mu v}{m^2_{\eta}+\frac{1}{2}(\lambda_4+\lambda_5+\lambda_8)v^2-(\mu^2_\sigma+\lambda_6 v^2)}~.
\end{align}
Under the assumption that the coupling $\mu$ is much smaller than the electroweak scale,
the diagonalized masses $\tilde{m}^2_{\eta_R}$ and $\tilde{m}^2_\sigma$ are approximately given by
\begin{align}
    \tilde{m}^2_{\eta_R}\simeq m^2_{\eta_R}~,\quad
    \tilde{m}^2_{\sigma}\simeq\mu^2_\sigma+\lambda_6 v^2~.
\end{align}

The mediator with an electroweak-scale mass is severely constrained by collider experiments.
In particular, the long-lived charged component of $\eta$ behaves similarly to a slepton and is constrained by searches for large ionization energy loss at colliders~\cite{ATLAS:2022pib,CMS:2024nhn,ATLAS:2025fdm}.
The CMS collaboration gave the bound on the left-handed stau as $m_{\tilde{\tau}_L} \gtrsim 660$\,GeV~\cite{CMS:2024nhn}.
We therefore restrict our analysis to mediator masses above
700~GeV throughout this paper.

\subsection{Lifetime}
\label{sec:lifetime}

\begin{figure}[tb]
\centering
\includegraphics[width=1.0\textwidth]{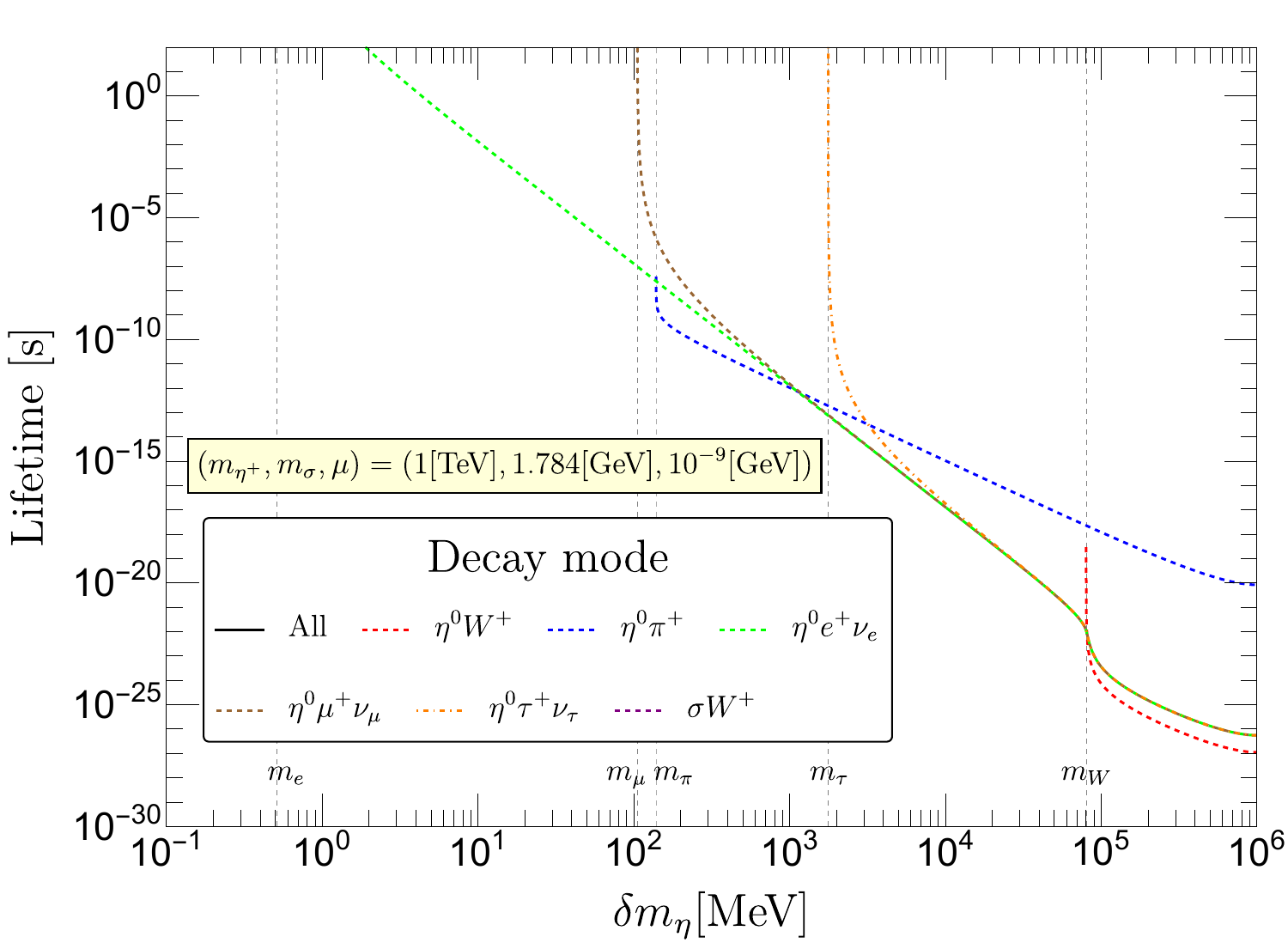}
\caption{Lifetime of the charged scalar $\eta^+$ as a function of the mass difference.}
\label{fig:lifetime}
\end{figure}

In Fig.~\ref{fig:lifetime}, the lifetime of the charged scalar, $\eta^+$, is shown as a black solid curve as a function of the mass difference $\delta m_\eta \equiv m_{\eta^\pm}-\tilde{m}_{\eta_R}$.
Colored dotted curves represent the time scale given by the inverse of the partial widths for each process.
The masses of $\eta^+$, $\sigma$ and coupling $\mu$ 
are taken with benchmark parameters for this scenario 
$m_{\eta^\pm} = 1$\,TeV, $m_\sigma = 1.784\,$GeV, and $\mu = 10^{-9}\,$GeV, respectively.
The vertical gray dashed lines indicate the thresholds at which new decay channels open, corresponding to the values of $\delta m_\eta$ matching the masses of SM particles.

As shown in Fig.~\ref{fig:lifetime}, the lifetime of $\eta^+$ with $\delta m_\eta\ge10$\,MeV is shorter than 0.01\,s regardless of the scalar trilinear coupling $\mu$, which controls the partial decay width of $\eta^+ \to \sigma W^+$.
Therefore, in the region $\delta m_\eta \gtrsim 10\,$MeV, the decay process $\eta^+ \to \eta^0 W^+$ does not affect the BBN, and we do not consider the lifetime of $\eta^+$ with $\delta m_\eta \gtrsim 10$\,MeV.
Hereafter, we focus on the case $\delta m_\eta\leq10$\,MeV.

\subsection{Cosmological aspects}
In the asymmetric mediator scenario, the lepton and mediator ($\eta$) asymmetries, defined as $n_{\Delta \text{L}} \equiv n_\text{L} - n_{\bar{\text{L}}}$ and $n_{\Delta \eta} \equiv n_{\eta^*}-n_{\eta}$, respectively, are generated via the CP-violating decays of heavy Majorana neutrinos.
The two asymmetries are equal such that $n_{\Delta \text{L}}=n_{\Delta\eta}$ immediately after the decay of the Majorana neutrino. 
However, the $\eta$-asymmetry may be washed out through the process $\eta\eta\leftrightarrow HH$ in a later stage.
To maintain the relation $n_{\Delta \text{L}}=n_{\Delta \eta}$, we focus the range
\begin{equation}
    \lambda_8 \lesssim 3.9\times 10^{-8} \sqrt{\frac{m_\eta}{\rm GeV}}~,
\end{equation}
to realize $\Gamma_{\eta\eta\leftrightarrow HH}(T)<H(T)$ in any temperature $T$, where $\Gamma_{\eta\eta\leftrightarrow HH}(T)$ and $H(T)$ denote the reaction rate and the Hubble parameter, respectively~\cite{Asai:2022vat}.
Once the asymmetries are established, the mediator $\eta$ and anti-mediator $\eta^*$ undergo pair annihilation via gauge interactions, such as $\eta \eta^* \to BB$ and $WW$, depleting the symmetric component and leaving only the asymmetric part.
The mass of the mediator $\eta$ must be smaller than around 100 TeV for the complete annihilation of the symmetric component, since the annihilation cross section becomes smaller for a higher mass of $\eta$.
After the $\eta$ pair annihilation,
the remaining asymmetric component of the mediator decays into the DM particle $\sigma$ and the SM particle.
As a result, the DM number density $n_\sigma$ matches the mediator asymmetry $n_{\Delta \eta}$, leading to a coincidence between the lepton asymmetry and DM abundance: $n_{\Delta \text{L}} = n_\sigma$.
The sufficiently small coupling $\mu$ is required for the late decay of $\eta$ after the complete annihilation, although $\mu$ plays an important role in the dark matter production and should not be zero.

The mass of the DM particle is determined by the cosmological density parameters $\Omega_{\rm DM}$, $\Omega_\text{B}$, and the relation $n_{\Delta \text{L}}=n_{\Delta \eta}$.
After the leptogenesis epoch, the generated lepton asymmetry $n_{\Delta \text{L}}$ is partially converted into the baryon asymmetry, given by $n_{\Delta \text{B}}=-28n_{\Delta \text{L}}/79$, through the sphaleron process during the electroweak phase transition~\cite{Harvey:1990qw}.
Since the DM candidate is identified with $\tilde{\sigma}$, the DM mass is determined by
\begin{align}
    m_{\text{DM}}=\tilde{m}_\sigma=\frac{28}{79}\frac{\Omega_{\text{DM}}}{\Omega_\text{B}}m_\text{B}=1.784\,\text{GeV},\label{DMmass}
\end{align}
where $\Omega_{\text{DM}}h^2=0.1200$, $\Omega_\text{B} h^2=0.02237$ measured by the Planck observations of cosmic microwave background~\cite{Planck:2018vyg}, and $m_\text{B}=938.3$ MeV for the proton mass \cite{ParticleDataGroup:2024cfk}. 
In this paper, we consider the case $\delta m_\eta>0$, i.e., the neutral particles $\eta_{R,I}$ are lighter than the charged particles $\eta^\pm$.
The case for $\delta m_\eta < 0$ can be interpreted by switching the roles of the charged and neutral particles.

\section{Relic abundances in scotogenic model}
\label{sec:pheno}
In this section, we evaluate the yields of $\eta$ and $\sigma$ particles during the BBN era by solving the Boltzmann equations, which include the decay and scattering processes.
See Refs.~\cite{Kolb:1990vq, Enomoto:2023cun} for detailed techniques concerning the Boltzmann equations.
Since the massive $\eta$ that decays into light particles spoils the successful BBN scenario, the model parameters are constrained.

\subsection{Decay processes}
\label{sec:decay}
Since we consider $\delta m_\eta = m_{\eta^+}-m_{\eta_R}>0$ and $\tilde{m}_{\eta_R}, m_{\eta_I}, m_{\eta^+} > m_h, m_Z, m_W \gg\tilde{m}_\sigma$ where $m_h, m_Z, m_W$ are the masses of the Higgs, $Z$, and $W$ bosons respectively, there are four significant decay channels of $\eta^\pm$ and two decay channels of $\eta_{R,I}$ as
\begin{itemize}
    \item $\eta^\pm \leftrightarrow \eta^0 W^\pm,\quad
    \eta^\pm \leftrightarrow \eta^0\pi^\pm,\quad
    \eta^\pm \leftrightarrow \eta^0 \ell^\pm \nu_\ell \quad (\ell=e,\mu,\tau),\quad
    \eta^\pm \leftrightarrow \tilde{\sigma}W^\pm$,
    \item $\eta_R \leftrightarrow \tilde{\sigma}h,\quad
    \eta_I \leftrightarrow \tilde{\sigma}Z$.
\end{itemize}
For $\eta^\pm \leftrightarrow \eta^0\pi^\pm$, we follow the result in Ref.~\cite{Avila:2021mwg}. Furthermore, we ignore the processes of $\eta^\pm \leftrightarrow \sigma\pi^\pm$, $\eta^\pm \leftrightarrow \sigma \ell^\pm \nu_\ell$, $\tilde{\eta}_R\leftrightarrow\tilde{\sigma}\pi^0$, and $\eta_I\leftrightarrow\tilde{\sigma}\pi^0$ since they are subdominant and negligible compared to the listed processes above: $\eta^\pm\leftrightarrow\sigma W^\pm$, $\tilde{\eta}_R\leftrightarrow\tilde{\sigma}h$, and $\eta_I\leftrightarrow\sigma Z$.
See Appendix \ref{app:decay} for the detailed calculation of each decay rate.

\subsection{Scattering processes}
\label{sec:scattering}
There are three dominant scattering processes:
\begin{itemize}
    \item $\eta\eta^*\leftrightarrow GG', \quad \eta^\pm\ell^\mp\leftrightarrow\eta^0\nu_\ell$, \\
    \item $\eta_{R,I} t\leftrightarrow\tilde{\sigma} t$,\quad $\eta_{R,I} \bar{t}\leftrightarrow\tilde{\sigma} \bar{t}$
\end{itemize}
where we denote $G,G'$ as the SM gauge bosons, $\ell$ as the charged leptons ($e,\mu,\tau$), and $t$ as the top quark.
See Appendix \ref{app:scattering} for the evaluation of the cross sections and their thermal averaging for each process.

In addition to the above processes, an $\eta$-number violating process $\eta\eta\leftrightarrow HH$ is likely viable. However, this must be negligible in our scenario to avoid the wash-out of the mediator asymmetry as mentioned in Sec.~\ref{sec:model}.

\subsection{Boltzmann equations}
\label{sec:Boltzmann}
We define the difference of the number density between $\eta$ and $\eta^*$ as $n_{\Delta\eta}\equiv n_{\eta^*}-n_{\eta}$ with $n_\eta\equiv n_{0}+n_{+}$ and $n_{\eta^*}\equiv n_{0^*}+n_{-}$.
Here, $n_\pm$ and $n_0,n_{0*}$ denote the number density of $\eta^\pm$, and $\eta^0,\eta^{0*}$, respectively.
The five kinds of Boltzmann equations concerning $\eta^+, \eta^-, \eta^0,\eta^{0*}$, and $\sigma$ need to be solved in order to fix each abundance at BBN.
Defining the evolution parameter $z$ as $z=m_{\eta_R}/T$ and the yield as $Y_i=n_i/s$ ($i$ for each species) where $s$ is the entropy density, the Boltzmann equations are given by
\begin{align}
    z\frac{d Y_{+}}{d z} & =-\gamma_\eta\left(Y_{+}-\frac{Y_{\pm}^\textrm{EQ}}{Y_{0}^\textrm{EQ}} Y_{0}\right)-\mathcal{D}_{\eta^+\to\sigma W^+}\left(Y_{+}-\frac{Y_{\pm}^\textrm{EQ}}{Y_{\sigma}^\textrm{EQ}} Y_{\sigma}\right)\nonumber \\
    &\quad-\mathcal{S}_{\eta^+ \eta^- \rightarrow GG'}\left(Y_{+}Y_{-}-(Y_{ \pm}^\textrm{EQ})^2\right)-\mathcal{S}_{\eta^+ \eta^0 \rightarrow G^\pm G'}\left(Y_{+}Y_{0^*}-Y_{\pm}^\textrm{EQ}Y_{0}^\textrm{EQ}\right), \label{Boltzmann:y+} \\
    z\frac{d Y_{-}}{d z} & = -\gamma_\eta\left(Y_{-}-\frac{Y_{\pm}^\textrm{EQ}}{Y_{0}^\textrm{EQ}} Y_{0^*}\right) -\mathcal{D}_{\eta^-\rightarrow\sigma W^-}\left(Y_{-}-\frac{Y_{\pm}^\textrm{EQ}}{Y_{\sigma}^\textrm{EQ}} Y_{\sigma}\right) \nonumber \\
    &\quad-\mathcal{S}_{\eta^+ \eta^- \rightarrow GG'}\left(Y_{+}Y_{-}-(Y_{ \pm}^\textrm{EQ})^2\right)-\mathcal{S}_{\eta^- \eta^0 \rightarrow G^- G'}\left(Y_{-}Y_{0}-Y_{\pm}^\textrm{EQ}Y_{0}^\textrm{EQ}\right), \label{Boltzmann:y-} \\
    z\frac{d Y_{0}}{d z} & =+\gamma_\eta\left(Y_{+}-\frac{Y_{ \pm}^\textrm{EQ}}{Y_{0}^\textrm{EQ}} Y_{0}\right)
    -\gamma_\sigma\left(Y_{0}-\frac{Y_{0}^\textrm{EQ}}{Y_{\sigma}^\textrm{EQ}} Y_{\sigma}\right) \nonumber \\
    &\quad-\mathcal{S}_{\eta^0 \eta^0 \rightarrow GG'}\left(Y_{0}Y_{0^*}-(Y_{0}^\textrm{EQ})^2\right)-\mathcal{S}_{\eta^- \eta^0 \rightarrow G^- G'}\left(Y_{-}Y_0-Y_{ \pm}^\textrm{EQ}Y_{0}^\textrm{EQ}\right), \label{Boltzmann:y0}\\
    z\frac{d Y_{0^*}}{d z} & =+\gamma_\eta\left(Y_{-}-\frac{Y_{ \pm}^\textrm{EQ}}{Y_{0}^\textrm{EQ}} Y_{0^*}\right)
    -\gamma_\sigma\left(Y_{0^*}-\frac{Y_{0}^\textrm{EQ}}{Y_{\sigma}^\textrm{EQ}} Y_{\sigma}\right) \nonumber \\
    &\quad-\mathcal{S}_{\eta^0 \eta^0 \rightarrow GG'}\left(Y_{0}Y_{0^*}-(Y_{0}^\textrm{EQ})^2\right) -\mathcal{S}_{\eta^+ \eta^0 \rightarrow G^+ G'}\left(Y_{+}Y_{0^*}-Y_{ \pm}^\textrm{EQ}Y_{0}^\textrm{EQ}\right), \label{Boltzmann:y0*}\\
    z\frac{d Y_{\sigma}}{d z} & =+\gamma_\sigma\left(Y_{0}+Y_{0*}-2\frac{Y_{0}^\textrm{EQ}}{Y_{\sigma}^\textrm{EQ}} Y_{\sigma}\right) +\mathcal{D}_{\eta^\pm\rightarrow\sigma W^\pm}\left(Y_{+}+Y_{-}-2\frac{Y_{ \pm}^\textrm{EQ}}{Y_{\sigma}^\textrm{EQ}} Y_{\sigma}\right), \label{Boltzmann:ysigma}
\end{align}
where we denote
\begin{align}
    \gamma_\eta & =\sum_\ell\left( \mathcal{D}_{\eta^+\to\eta^0\ell^+\nu_\ell}+\mathcal{S}_{\eta^+\ell^-\to\eta^0\nu_\ell}Y_\ell^\textrm{EQ}+\mathcal{S}_{\eta^+\nu_\ell\to\eta^0\ell^-}Y_{\nu_\ell}^\textrm{EQ}\right), \\
    \gamma_\sigma & = \mathcal{D}_{\eta^0\to\sigma h}+\mathcal{D}_{\eta^0\to\sigma Z}+\mathcal{S}_{\eta^0t\to\sigma t}Y_t^\textrm{EQ}, \\
\end{align}
and $\mathcal{D}_X$, $\mathcal{S}_X$ ($X$ for each process) are functions of $z$ describing the thermally-averaged reaction rate normalized by the Hubble parameter, defined by
\begin{align}
    \mathcal{D}_X & = \left(1+\frac{1}{3}\frac{d\ln h}{d\ln T}\right)\frac{\langle\Gamma_X\rangle}{H}, \\
    \mathcal{S}_X & = \left(1+\frac{1}{3}\frac{d\ln h}{d\ln T}\right)\frac{s\langle\sigma_X v\rangle}{H},
\end{align}
respectively
\footnote{
The factor $1+\frac{1}{3}\frac{d\ln h}{d\ln T}$ originates from the variable transformation of time $t$ to the evolution parameter $z$.
For more detail, see Refs.~\cite{Gondolo:1990dk,Drees:2015exa,Saikawa:2020swg}.
}.
Here, the Hubble parameter $H$, the entropy density $s$, and the yield in thermal equilibrium $Y_i^\textrm{EQ}$ ($i$ for each species) are defined by
\begin{align}
    H
    =\sqrt{\frac{4 \pi^{3} g(T)}{45}} \frac{m^{2}_{\eta_R}}{m_{\rm Pl} z^{2}},\quad
    s
    =\frac{2 \pi^{2}}{45} \frac{m^{3}_{\eta_R} h(T)}{z^{3}},\quad
    Y^\textrm{EQ}_i=\frac{45}{4\pi^4}\frac{g_i}{h(T)}\left(\frac{m_i}{m_{\eta_R}}\right)^2z^2 K_2\left(\frac{m_i}{m_{\eta_R}}z\right),
\end{align}
where $m_{\rm Pl}=1.22\times 10^{19}$ GeV is the Planck mass, $g_i$ is the degrees of freedom (d.o.f.) for species $i$, and $g(T)$, $h(T)$ count the effective number of relativistic d.o.f. that contribute to the energy density and the entropy density, respectively. $K_2(x)$ is the modified Bessel function of the second kind.
Especially, the temperature dependence of $g(T)$ and $h(T)$ is referred to Ref.~\cite{Saikawa:2020swg} in our analysis.

Since the asymmetry between $\eta$ and $\eta^*$ exists, we must pay attention to the initial conditions for the Boltzmann equations.
The component of the asymmetry of $\eta$ is given by
\begin{align}
    Y_{\Delta \eta} & = Y_{\eta^*}-Y_\eta = (Y_{0^*}+Y_-)-(Y_0+Y_+)~.
\end{align}
Assuming that the asymmetry $Y_{\eta^*}-Y_\eta$ is equally distributed between $Y_{0^*}-Y_0$ and $Y_--Y_+$, we set the initial yields as
\begin{align}
    Y_{+}(z_i)&=Y^\textrm{EQ}_{\pm}(z_i)-\frac{1}{4}Y_{\Delta \eta}~, \quad
    Y_{-}(z_i)=Y^\textrm{EQ}_{\pm}(z_i)+\frac{1}{4}Y_{\Delta \eta}~, \\
    Y_0(z_i)&=Y^\textrm{EQ}_{0}(z_i)-\frac{1}{4}Y_{\Delta \eta}~, \quad
    Y_{0^*}(z_i)=Y^\textrm{EQ}_{0^*}(z_i)+\frac{1}{4}Y_{\Delta \eta}~,
\end{align}
where $z_i$ is an initial value of $z$. The value of $Y_{\Delta\eta}$ can be estimated from the present baryon asymmetry:
\begin{equation}
    Y_{\Delta \eta}(z_i) = -Y_{\Delta \text{L}}(z_i) = \frac{79}{28}Y_{\Delta \text{B}}^{\rm obs} = 2.44\times 10^{-10},
\end{equation}
where we used the observed baryon number $Y_{\Delta \text{B}}^{\rm obs}=8.65\times 10^{-11}$. We also set the initial condition for $\sigma$ as $Y_{\sigma}(z_i)=0$ because $\sigma$ is out of thermal equilibrium at the initial.

\section{Numerical results}
\label{sec:result}

\subsection{Relic abundance}
\label{subsec:relic}

\begin{figure}[tbp]
\begin{minipage}[b]{0.5\linewidth}
\centering
\includegraphics[keepaspectratio, scale=0.6]{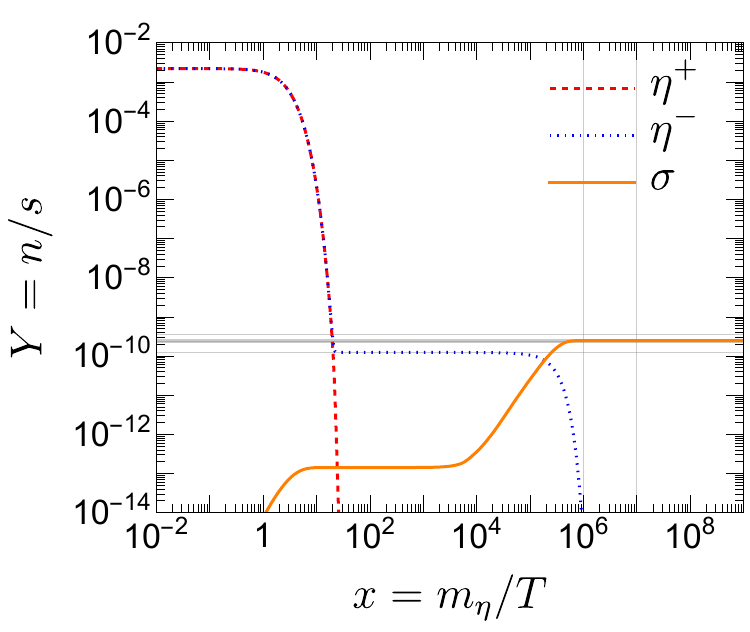}
\subcaption{Charged $\eta$}
\end{minipage}
\begin{minipage}[b]{0.5\linewidth}
 \centering
\includegraphics[keepaspectratio, scale=0.6]{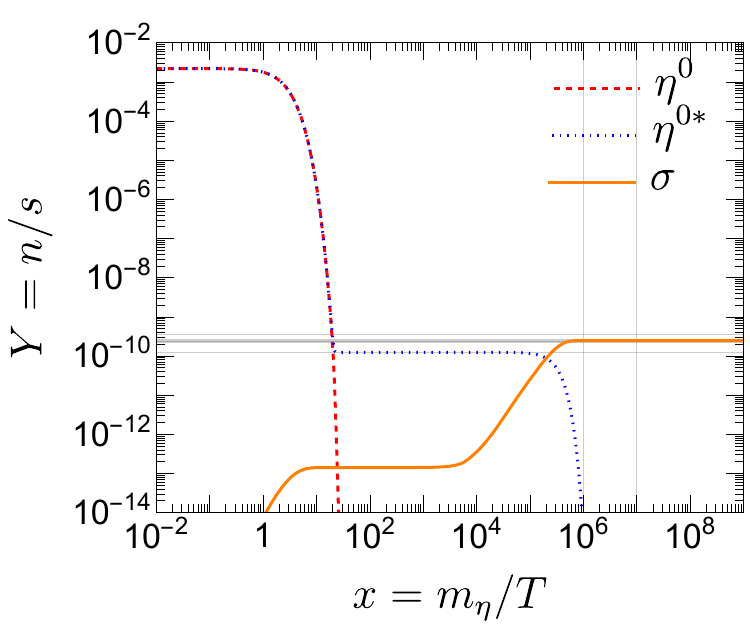}
\subcaption{Neutral $\eta$}
\end{minipage}
\caption{
Numerical results for the evolution of each yield.
The left (right) panel shows the evolution of the yields of the charged (neutral) mediators and the dark matter.
The dashed red, dotted blue, and solid orange lines in the left (right) panel correspond to 
$\eta^+$ ($\eta^0$), $\eta^-$ ($\eta^{0*}$), and $\sigma$, respectively.
The parameters are fixed as $m_{\eta_R}=1$\,TeV, $\lambda_8=6.3\times10^{-8}$, $\mu=10^{-9}$\,GeV, and $\delta m_\eta=1$\,MeV.
The horizontal gray lines show the yields of the lepton asymmetry generated via thermal leptogenesis.
The vertical gray lines indicate the cosmic temperature at 1\,MeV, corresponding to the onset of BBN.
}
\label{fig:Boltzmanneq}
\end{figure}

Figure~\ref{fig:Boltzmanneq} shows the results for solving the Boltzmann equations \eqref{Boltzmann:y+}-\eqref{Boltzmann:ysigma} with parameters $\tilde{m}_{\eta_R}=1$\,TeV, $\lambda_8=6.3\times10^{-8}$, $\mu=10^{-9}$\,GeV, and $\delta m_\eta=1$\,MeV.
In the left (right) panel of Fig.~\ref{fig:Boltzmanneq}, the red and blue lines represent the yields of $\eta^+$ ($\eta^0$) and $\eta^{-}$ ($\eta^{0*}$), respectively.
The orange line in both panels shows the yield of $\sigma$.
The horizontal gray solid line shows the yield of the lepton asymmetry generated via the leptogenesis scenario to explain BAU, given as $Y_{\Delta \text{L}} = (79/28) Y_{\Delta \text{B}}^{\rm obs} = 2.44 \times  10^{-10}$. 
The vertical gray solid line labeled ``BBN'' indicates $T=1$ MeV, which corresponds to one second in cosmic time approximately.
As shown in Fig.~\ref{fig:Boltzmanneq}, the yield of $\sigma$ (DM) reaches $Y_\sigma = \epsilon$ as the temperature decreases.

\begin{figure}[tbp]
\begin{minipage}[b]{0.5\linewidth}
\centering
\includegraphics[keepaspectratio, scale=0.6]{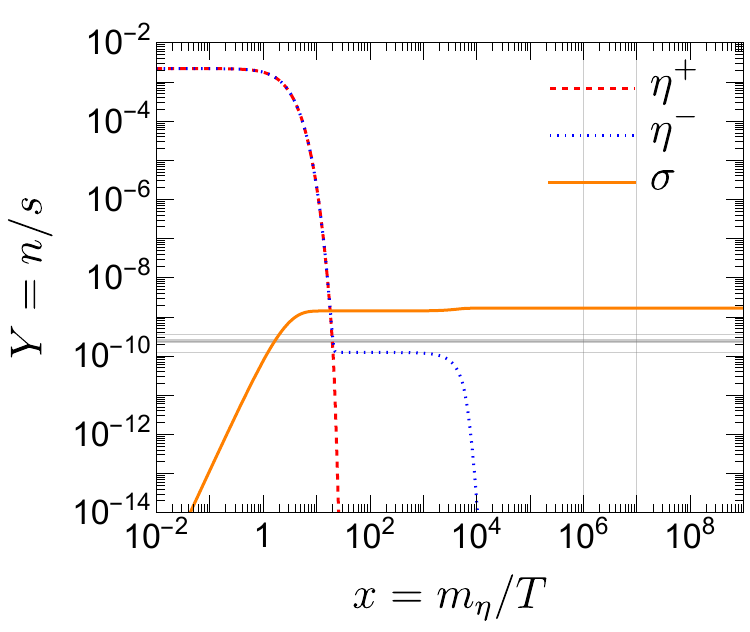}
\subcaption{Charged $\eta$}
\end{minipage}
\begin{minipage}[b]{0.5\linewidth}
 \centering
\includegraphics[keepaspectratio, scale=0.6]{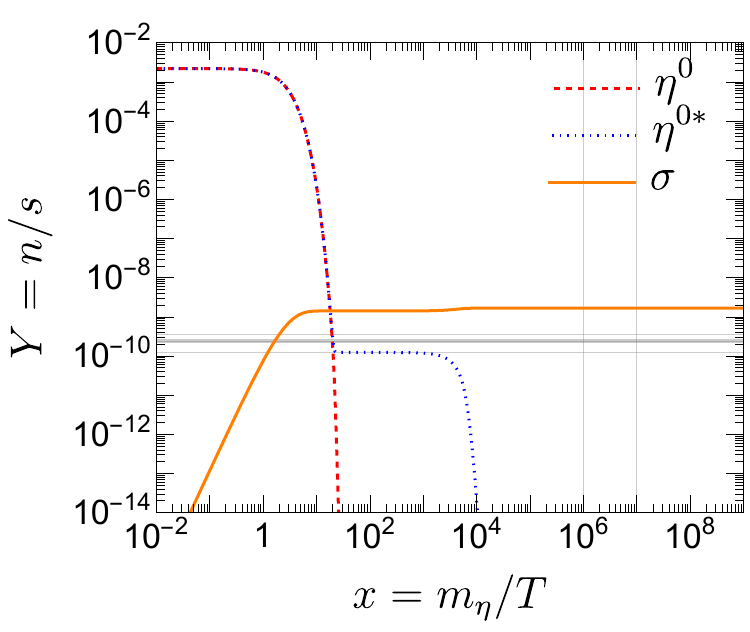}
\subcaption{Neutral $\eta$}
\end{minipage}
\caption{
Same as Fig.~\ref{fig:Boltzmanneq}, but for $\mu=10^{-7}$\,GeV.
}
\label{fig:Boltzmanneq2}
\end{figure}

Figure~\ref{fig:Boltzmanneq2} shows the similar results to those in Fig.~\ref{fig:Boltzmanneq}, except for a different value of $\mu=10^{-7}$\, GeV instead of $\mu=10^{-9}$\, GeV.
In this case, the yield of the DM exceeds the value of $Y_{\Delta \text{L}}$; that is, DM particles are overproduced. 
This overproduction arises from the decays of the mediator driven by the trilinear coupling $\mu$ as $\eta^\pm \to \sigma W^\pm$ and $\eta^0 \to \sigma h/\sigma Z$, like the freeze-in mechanism before the annihilation of the mediators. 
Therefore, the parameter region that leads to DM overproduction must be excluded.

Figure~\ref{fig:BBN:1MeV} shows the allowed parameter region 
in the $m_\eta$–$\mu$ plane that explains the observed DM relic abundance.
The regions below the solid red, dashed orange, and dotted blue lines correspond to $r\equiv(Y_\sigma - Y_{\Delta \text{L}}) / Y_{\Delta \text{L}} \leq 0.01$, $0.1$, and $1$, respectively.
As shown in Fig.~\ref{fig:BBN:1MeV}, the region with 
heavier $m_\eta$, e.g., $m_\eta \gtrsim$ $2 \times 10^5$\,GeV for $r=1$, predicts the overabundance of the DM relic density, regardless of the size of the scalar trilinear coupling $\mu$.
This is because the annihilation cross section of the heavier mediators becomes too small to efficiently eliminate their symmetric component of the mediators, resulting in excess DM production by the subsequent decays of the remaining mediators.
Moreover, the boundary lines slope upward up to around $10^5$\, GeV.
This behavior originates from the generation of $\sigma$ particles via the freeze-in mechanism, and is described as
$\mu\propto m_\eta^{3/2}$.  See Appendix~\ref{app:production} in detail.
Therefore, if the mediator mass is sufficiently large, the decay of the mediators occurs significantly later than their annihilation, even though their lifetimes become shorter for larger values of $\mu$.

\subsection{Comparison to BBN constraints}
\label{subsec:BBN}

Before discussing the allowed region of our model, we briefly introduce the BBN constraints~\cite{Kawasaki:2004qu}.
Figures 38-40 in Ref.~\cite{Kawasaki:2004qu}
give the constraint on the lifetime of the long-lived particle $X$ ($\eta$ in this paper) and the yield of it.

The constraints from BBN are obtained from three processes, i.e., proton-neutron interconversion, hadrodissociation, and photodissociation processes.
The proton-neutron interconversion process ($p \leftrightarrow n$) is induced by hadrons emitted from the decay of the long-lived particle $X$.
The emitted hadrons, such as charged pions ($\pi^\pm$), convert ambient protons and neutrons through strong interactions, $n+\pi^+\to p+\pi^0$ and $p+\pi^-\to n+\pi^0$ in the early stage of the BBN.
As a result, the neutron-to-proton ratio ($n/p$) tends to approach unity, leading to the overproduction of ${}^4$He.
Therefore, the long-lived particle with short lifetime ($\tau_X \lesssim 100$\,s) is constrained by the proton-neutron interconversion.
In the latter stage of the BBN ($\tau_X \gtrsim 100$\,s), energetic hadrons produced by the decay of $X$ break light-elements into pieces and change their abundances.
The third process is the photodissociation.
High-energy charged leptons and photons produced by the decay of $X$ react with the background photons and then lose their energy.
The low-energy photons, whose energy fall below the threshold of electron-positron pair production with the background photons, can destroy the light elements.
The photodissociation processes of the deuterium D and helium ${}^4$He become effective in the temperatures lower than $T \lesssim 0.01$\,MeV and $T \lesssim 0.001$\,MeV, respectively.
Therefore, the constraint on $X$ from the photodissociation process is significant in $t \lesssim 10^{4-6}$\,s.

According to the Figures 38-40 in Ref.~\cite{Kawasaki:2004qu}, the most stringent constraint on $X$ for the lifetime $\tau\le100$\,s comes from the ${}^4$He mass fraction.
There are two bounds derived from ${}^4$He, denoted as $Y_p$(IT) and $Y_p$(FO), which reflect differences in the inferred values of the primordial helium abundance reported in Ref.~\cite{Fields:1998gv,Izotov:2003xn} as follows~:
\begin{align*}
   Y_p^{\rm obs}({\rm FO}) &= 0.238 \pm (0.002)_{\rm stat} \pm (0.005)_{\rm syst}~, \\
   Y_p^{\rm obs}({\rm IT}) &= 0.242 \pm (0.002)_{\rm stat} \left(\pm (0.005)_{\rm syst} \right)~.
\end{align*}
The recent analysis of observational data from the extremely metal-poor galaxies yields the following result~\cite{Matsumoto:2022tlr}
\begin{align*}
   Y_p^{\rm obs}({\rm EMPRESS}) = 0.2370_{-0.0033}^{+0.0034}~,
\end{align*}
which is close to the value reported in Ref.~\cite{Fields:1998gv}.
Therefore, we consider that it is more appropriate to apply the bound based on $Y_p^{\rm obs}({\rm FO})$.
For the lifetime $\tau \gtrsim 100$\,s, the most stringent constraints on $X$ come from the fractions D/H, ${^6}$Li/H, and ${}^3$He/D.

In Ref.~\cite{Kawasaki:2004qu}, the constraints on $X$ are given in the plane of the lifetime $\tau_X$ and $E_\textrm{vis} Y_X$ with the yield of $X$ at the temperature $T = 1$\,MeV, $Y_X$, and visible energy released by the single decay of $X$, $E_\textrm{vis}$.
In this paper, the visible energy is a half of mediator mass as $E_\textrm{vis} = m_\eta/2$, since we focus on the case for $\delta m_\eta \le 10$\,MeV, and the charged (neutral) mediators dominantly decay into pairs of the dark matters and $W$ bosons (Higgs bosons).

For the recast the constraints in Ref.~\cite{Kawasaki:2004qu} to apply them into arbitrary mass and hadronic branching ratio of the mediator $\eta$, we define the equality-time $t_\textrm{eq}$.
The equality-time $t_\textrm{eq}$ is the time when the contribution to the light element destruction from the hadronic processes (proton-neutron interconversion and hadrodissociation) becomes equal to that from electromagnetic one (photodissociation) and can be picked up from the lifetimes when the bound lines in the Figures 38-40 in Ref.~\cite{Kawasaki:2004qu} are hollowed and become weaker.
In our recast, we choose the equality-times for each constraints as follows:
\begin{align}
   t_\textrm{eq}[Y_p\textrm{(FO)}] = 10^4\,\textrm{s}, \qquad
   t_\textrm{eq}[\textrm{D/H(low)}] = 10^8\,\textrm{s}, \qquad
   t_\textrm{eq}[{}^6\textrm{Li/H}] = 10^8\,\textrm{s}~.
\end{align}

After the equality-time, the contribution from the photodissociation process is dominant, and the strength of the bounds is proportional to the energy emitted from a decay of the mediator.
Therefore, the upper bound on $E_\textrm{vis} Y_X$ for $t > t_\textrm{eq}$ in our model can be recast by using that for $m_X = 1$\,TeV and $B_h=1$ in Kawasaki-Kohri-Moroi's (KKM) work~\cite{Kawasaki:2004qu}, as
\begin{align}
   (Y_\eta)_\textrm{our}
   <  2 \times 10^{-3} \left( \frac{m_\eta}{1\,\textrm{TeV}} \right)^{-1} \frac{(E_\textrm{vis} Y_X)_\textrm{KKM}^\textrm{upper bound}}{\textrm{GeV}}~,
\end{align}
with $Y_\eta \equiv Y_{\eta^+} + Y_{\eta^-} + Y_{\eta^0} + Y_{\eta^{0*}}$.

On the other hand, the contributions from the hadronic processes are dominant before the equality-time and then depend on the hadronic branching ratio of the mediator.
The strength of the bounds from the hadronic processes are proportional to the number of hadrons produced through the hadronization process, which is proportional to $(m_\eta)^\delta$ with $\delta \sim 0.3$ when all the decay modes are hadronic~\cite{Kawasaki:2017bqm}.
The energy converted into the hadron sector is calculated as
\begin{align}
   \textrm{E}_\textrm{had}^{\eta^\pm} &= 
   \frac{m_\eta}{2} \sum_{X= \eta^0, \tilde{\sigma}} \left[ \textrm{Br}(\eta^\pm \to X + \textrm{had}) \right. \nonumber \\
   &\left. \hspace{18mm} + \textrm{Br}(\eta^\pm \to X W^\pm) \left\{ \textrm{Br}(W^\pm \to \textrm{had}) + \textrm{Br}(W^\pm \to \tau^\pm \nu_\tau) \, \textrm{Br}(\tau^\pm \to \nu_\tau + \textrm{had})/4 \right\} \right] \nonumber \\
   &\simeq \frac{m_{\eta}}{2} \textrm{Br}(\eta^\pm \to \tilde{\sigma} W^\pm) \left\{ \textrm{Br}(W^\pm \to \textrm{had}) + \textrm{Br}(W^\pm \to \tau^\pm \nu_\tau) \, \textrm{Br}(\tau^\pm \to \textrm{had})/4 \right\} \nonumber \\
   &= 0.35 \, m_\eta \times \textrm{Br}(\eta^\pm \to \tilde{\sigma} W^\pm)~, \\
   \textrm{E}_\textrm{had}^{\eta^0} &= 
   \frac{m_\eta}{2} \, \textrm{Br}(\eta^0 \to \sigma h) \left[ \textrm{Br}(h \to \textrm{had}) + \textrm{Br}(h \to \tau^+ \tau^-) \right. \nonumber \\
   &\left. \hspace{38mm} \times \left\{ \textrm{Br}(\tau^\pm \to \nu_\tau + \textrm{had})^2/2 + \textrm{Br}(\tau^\pm \to \nu_\tau + \textrm{had}) \textrm{Br}(\tau^\pm \to \textrm{leptons})/4 \right\} \right] \nonumber \\
   & \quad +\frac{m_\eta}{2} \, \textrm{Br}(\eta^0 \to \sigma Z) \left[ \textrm{Br}(Z \to \textrm{had}) + \textrm{Br}(Z \to \tau^+ \tau^-) \right. \nonumber \\
   &\left. \hspace{38mm} \times \left\{ \textrm{Br}(\tau^\pm \to \nu_\tau + \textrm{had})^2/2 + \textrm{Br}(\tau^\pm \to \nu_\tau + \textrm{had}) \textrm{Br}(\tau^\pm \to \textrm{leptons})/4 \right\} \right] \nonumber \\
   &= 0.37 \, m_\eta  \times \textrm{Br}(\eta^0 \to \sigma h)+0.35 \, m_\eta  \times \textrm{Br}(\eta^0 \to \sigma Z)~.
\end{align}
We focus on the mass region of $\delta m_\eta$ as $\delta m_\eta < 10$\,MeV, and thus the decay channel of $\eta^\pm \to \eta^0 + (\textrm{had})$ is not taken into account.
Therefore, the upper bound on $Y_\eta$ for $t < t_\textrm{eq}$ in our model can be recast by using that for $m_X=1$\, and $B_h=1$ in Ref.~\cite{Kawasaki:2004qu}, $(E_\textrm{vis} Y_X)_\textrm{KKM}^\textrm{bound}$, as
\begin{align}
   (Y_\eta)_\textrm{our}^\textrm{upper bound}
   = 2 \times 10^{-3} \left( \frac{m_\eta}{1\,\textrm{TeV}} \right)^{-1} \left( \frac{E_\textrm{had}^\eta}{1\,\textrm{TeV}} \right)^{-\delta} \frac{(E_\textrm{vis} Y_X)_\textrm{KKM}^\textrm{upper bound}}{\textrm{GeV}}~. 
\end{align}

\begin{figure}[tbp]
\begin{minipage}[b]{0.5\linewidth}
\centering
\includegraphics[keepaspectratio, scale=0.63]{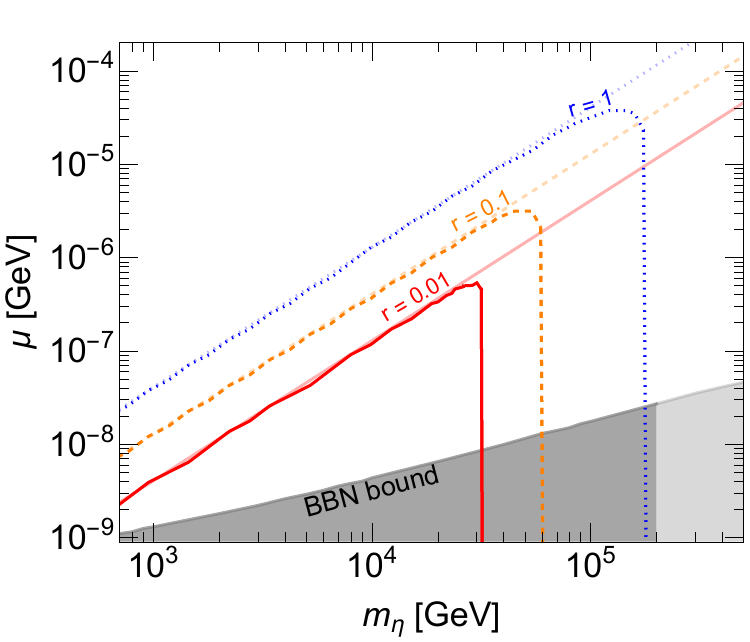}
\subcaption{$\delta m_\eta=1$\,MeV}
\end{minipage}
\begin{minipage}[b]{0.5\linewidth}
 \centering
\includegraphics[keepaspectratio, scale=0.63]{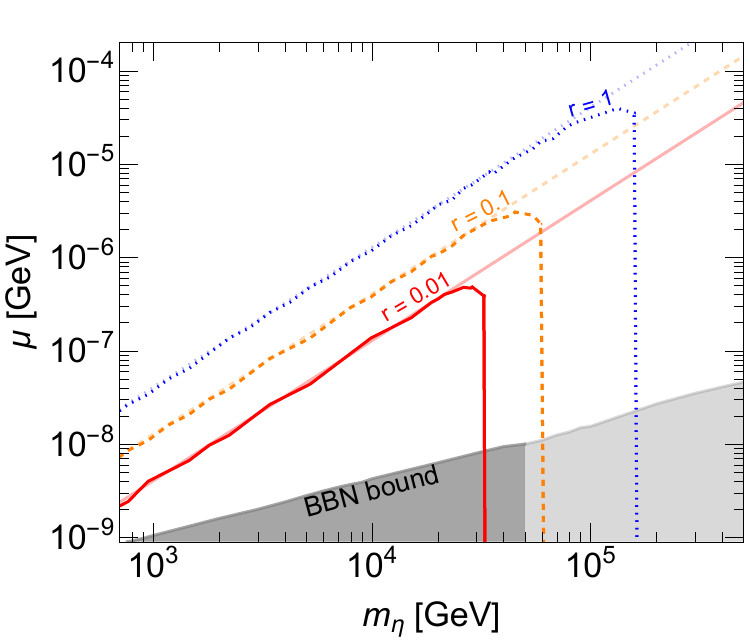}
\subcaption{$\delta m_\eta=10$\,MeV}
\end{minipage}
\caption{
BBN bound from Ref.~\cite{Kawasaki:2004qu} at $\delta m_\eta=1$\,MeV (left panel) and $\delta m_\eta=10$\,MeV (right panel).
The solid red, dashed orange, and dotted blue lines correspond to $r = 0.01, 0.1,$ and 1, respectively.
The light solid red, light dashed orange, and light dotted blue straight lines are the approximated results shown in Eq.~\eqref{eq:result-approx}. 
The dark and light gray shaded region is excluded by the BBN constraint based on Figure~39 of Ref.~\cite{Kawasaki:2004qu}, and the light one is obtained by extrapolating the upper bound on $Y_p$(FO) in Figure~39 of Ref.~\cite{Kawasaki:2004qu}.
}
\label{fig:BBN:1MeV}
\end{figure}

The results of recasting the upper bounds on the yield of the mediators for $\delta m_\eta = 1$\,MeV and $10$\,MeV are shown as dark gray shaded region in Fig.~\ref{fig:BBN:1MeV}.
The upper bound on the yield of $\eta$ comes from the constraint on $Y_p$ since the lifetime of the mediator is shorter than $10^4$\,s in the parameter region where we focus on.
The bound shown as light gray shaded region corresponds to the region of $E_\textrm{vis} Y_X > 10^{-6}$\,GeV in Figure~39 of Ref.~\cite{Kawasaki:2004qu}, and we obtain the upper bound on $Y_\eta$ by extrapolating the upper bound on $Y_p$(FO) in Figure~39 of Ref.~\cite{Kawasaki:2004qu}.
As shown in Fig.~\ref{fig:BBN:1MeV}, the smaller scalar coupling $\mu$ is excluded by the BBN bound since the mediators become so long lived that they decay during the BBN and change the abundance of the light elements.

\section{Summary}
\label{sec:summary}

The asymmetric mediator scenario provides novel insights into several long-standing 
puzzles in particle physics, including the origin of neutrino masses and mixing, 
as well as the remarkable closeness between the energy densities of dark matter 
(DM) and baryonic matter. 
Focusing on two cosmological issues, we investigated the favored and excluded 
parameter regions, aiming to test the viability of this scenario through terrestrial 
experiments and cosmological observations.

The first issue is an additional production channel for DM. 
The previous literature has considered DM production via the decay of the SU(2)$_L$ doublet field $\eta$, occurring only after the freeze-out of the $\eta$ asymmetry. 
Under this assumption, and given the common origin of the lepton and $\eta$ asymmetries, the energy densities of DM and baryonic matter are naturally connected. 
In contrast, the additional channels examined here take place prior to the freeze-out of the $\eta$ asymmetry, leading to qualitatively different implications for the resulting DM abundance.
To assess these effects quantitatively, we solved the network of Boltzmann equations of $\eta$ and $\sigma$ involving all relevant DM production channels. 
We derived a stringent upper bound on the interaction 
$\mu \sigma \left[ \left( H^\dagger \eta \right) + (\text{H.c.}) \right]$, 
which is required to prevent the overproduction of DM through processes such as
$\eta^\pm \to \sigma W^\pm$ and $\eta^0 \to \sigma h/\sigma Z$.

The second issue is non-standard nuclear reactions induced by long-lived $\eta$, which 
could spoil the successful BBN predictions. To avoid such effects, the $\eta$ lifetime 
must be short to ensure its decay occurs prior to the beginning of BBN era. 
The lifetime depends on $m_\eta$, the coupling $\mu$, and the mass splitting between 
the charged and neutral components $\delta m_\eta = m_{\eta^\pm} - \tilde{m}_{\eta_R}$. In scotogenic models, 
$\delta m_\eta \lesssim 100\,\text{MeV}$ arises naturally. Likewise, small values of $\mu$ 
and $m_\eta < \mathcal{O}(10)\,\text{TeV}$ are well motivated in the context of the 
asymmetric mediator scenario. 
To ensure the successful predictions of BBN, we updated the favored and excluded 
parameter regions in the $m_\eta$–$\mu$ plane for $\delta m_\eta = 1\,\text{MeV}$ and 
$10\,\text{MeV}$ as shown in Figure~\ref{fig:BBN:1MeV}.
The derived bounds play a crucial role in assessing the viability of the scenario 
in light of other theoretical and experimental considerations.

\section*{Acknowledgments}
We would like to thank Yuhei Sakai for useful discussion in the early stage of this study. 
This work was financially supported by JSPS KAKENHI Grant Numbers JP20H05852 [MY], 
JP22K03638 [MY], JP22K03602 [MY], JP23K13097 [KA], JP25KJ0401 [KA], and by the 
educational support fund from Kyushu Sangyo University [SE, TH].
The works of SE are also supported in part by the MEXT Leading Initiative for 
Excellent Young Researchers Grant Number 2023L0013. 
This work was partly supported by MEXT Joint Usage/Research Center on 
Mathematics and Theoretical Physics JPMXP0619217849 [MY].

\appendix
\section{Decay width}
\label{app:decay}
In this appendix, we summarize all results of decay width.
For $\eta^\pm$ decay, 
\begin{align}
    \Gamma(\eta^+ \to \eta^0 W^+)
    &= 
    \frac{g^2}{32 \pi m_{\eta^\pm}^3 m_W^2} \left\{ ( \delta m_\eta^2 - m_W^2 )( ( 2 m_{\eta^\pm} - \delta m_\eta )^2 - m_W^2 ) \right\}^\frac{3}{2}, \\
    \Gamma(\eta^+ \to \eta^0 \pi^+) &=
    2 \times \frac{f_\pi^2 g^4}{m_W^4} \frac{(m_{\eta^\pm}^2 - m_{\eta^0}^2)^2}{512 \pi m_{\eta^\pm}} \sqrt{1 - \left( \frac{m_{\eta^0} - m_\pi}{m_{\eta^\pm}} \right)^2} \sqrt{1 - \left( \frac{m_{\eta^0} + m_\pi}{m_{\eta^\pm}} \right)^2}, \\
    \Gamma(\eta^+ \to \eta^0 \ell^+ \nu_\ell)&= 
    \theta(\delta m_\eta - m_\ell) \cdot \frac{g^4}{512 \pi^3 m_{\eta^\pm}^3} \nonumber \\
    & \times \int_{m_\ell}^{\delta m_\eta} \dd P_{12} P_{12} \sqrt{\left( \frac{(2 m_{\eta^\pm} - \delta m_\eta)^2}{P_{12}} - P_{12} \right) \left( \frac{\delta m_\eta^2}{P_{12}} - P_{12} \right)} \nonumber \\
    & \quad \times \left( P_{12} - \frac{m_\ell^2}{P_{12}} \right) \cdot \frac{1}{\left( P_{12}^2 - m_W^2 \right)^2 + m_W^2 \Gamma_W^2} \nonumber \\
    & \quad \times \Bigg[ \left( \frac{\delta m_\eta (2 m_{\eta^\pm} - \delta m_\eta)}{P_{12}} + P_{12} \right)^2 \left( P_{12} + \frac{m_\ell^2}{P_{12}} \right)  \left( P_{12} - \frac{m_\ell^2}{P_{12}} \right) \nonumber \\
    & \qquad \quad + \frac{1}{3} \left( \frac{(2 m_{\eta^\pm} - \delta m_\eta)^2}{P_{12}} - P_{12} \right) \left( \frac{\delta m_\eta^2}{P_{12}} - P_{12} \right) \left( P_{12} - \frac{m_\ell^2}{P_{12}} \right)^2 \nonumber \\
    & \qquad \quad - 4 m_{\eta^\pm}^2 \left( P_{12}^2 - m_\ell^2 \right) \nonumber \\
    & \qquad \quad - 2 m_\ell^2 \left(1 + \frac{\delta m_\eta (2 m_{\eta^\pm} - \delta m_\eta)}{m_W^2} \right) \left( \frac{\delta m_\eta (2 m_{\eta^\pm} - \delta m_\eta)}{P_{12}} + P_{12} \right) \left( P_{12} - \frac{m_\ell^2}{P_{12}} \right) \nonumber \\
    & \qquad \quad + m_\ell^2 \left(1 + \frac{\delta m_\eta (2 m_{\eta^\pm} - \delta m_\eta)}{m_W^2} \right)^2 \left( P_{12}^2 - m_\ell^2 \right) \Bigg], \\
    \Gamma(\eta^+ \to \tilde{\sigma} W^+) &=
    \frac{g^2 \sin^2 \theta}{64 \pi m_{\eta^\pm}^3 m_W^2} \left\{ ( m_{\eta^\pm} + \tilde{m}_\sigma + m_W )( m_{\eta^\pm} + \tilde{m}_\sigma - m_W ) \right. \nonumber \\
    & \left. \hspace{25mm} \times ( m_{\eta^\pm} - \tilde{m}_\sigma + m_W )( m_{\eta^\pm} - \tilde{m}_\sigma - m_W ) \right\}^\frac{3}{2},
\end{align}
where $P_{12} = \sqrt{(p_{\eta^\pm} - p_{\eta^0})^2} = E_{\eta^\pm} - E_{\eta^0}$ in the off-shell $W^+$ rest frame, and $\delta_\eta \equiv (m_{\eta^\pm}^2 - m_{\eta^0}^2) / m_W^2$.
For $\eta_{R,I}$ decay,
\begin{align}
    \Gamma(\tilde{\eta}_R \to \tilde{\sigma} h) =&
    \frac{1}{16 \pi \tilde{m}_{\eta_R}^3}\left[ ( \tilde{m}_{\eta_R} + \tilde{m}_\sigma + m_h )( \tilde{m}_{\eta_R} + \tilde{m}_\sigma - m_h ) \right. \nonumber \\
    & \left. \hspace{25mm} \times ( \tilde{m}_{\eta_R} - \tilde{m}_\sigma + m_h )( \tilde{m}_{\eta_R} - \tilde{m}_\sigma - m_h ) \right]^\frac{1}{2} \nonumber \\
    & \times \left[(\lambda_4+\lambda_5-2\lambda_6+\lambda_8)v\sin\theta\cos\theta-\frac{1}{\sqrt{2}}\mu(\cos^2\theta-\sin^2\theta) \right]^2, \label{etaRtosigmah} \\
    \Gamma(\eta_I \to \tilde{\sigma} Z) =&
    \frac{(g^2+g'^2) \sin^2 \theta}{64 \pi m_{\eta_I}^3 m_Z^2} \left\{ ( m_{\eta_I} + \tilde{m}_\sigma + m_Z )( m_{\eta_I} + \tilde{m}_\sigma - m_Z ) \right. \nonumber \\
    & \left. \hspace{25mm} \times ( m_{\eta_I} - \tilde{m}_\sigma + m_Z )( m_{\eta_I} - \tilde{m}_\sigma - m_Z ) \right\}^\frac{3}{2}.
\end{align}
Translating into the component $\eta^0$, each decay rate becomes $\Gamma(\eta^0\to\sigma H)=\Gamma(\tilde{\eta}_R \to \tilde{\sigma} h)/2$, $\Gamma(\eta^0\to\sigma Z)=\Gamma(\tilde{\eta}_I \to \tilde{\sigma} Z)/2$.

\section{Scattering cross section of mediator}
\label{app:scattering}
In this appendix, we summarize the scattering cross sections.
For $\eta^+\ell^-\rightarrow\eta^0\nu_\ell$, the thermal averaged product of the scattering cross section and relative velocity $\langle\sigma v\rangle_{\eta^+\ell^-\rightarrow\eta^0\nu_\ell}$ is obtained as
\begin{align}
    &\langle\sigma v\rangle_{\eta^+\ell^-\rightarrow\eta^0\nu_\ell} \nonumber \\
    =&\frac{1}{16m^2_\pm m^2_\ell T K_2(m_\pm/T) K_2(m_\ell/T)}\int_{(m_{\eta^\pm}+m_\ell)^2}^\infty \frac{ds}{\sqrt{s}}K_1\left(\frac{\sqrt{s}}{T}\right) \theta(\sqrt{s} - m_{\eta^0}) \cdot \frac{1}{32 \pi}\cdot 2 \cdot \frac{g^4}{4} \nonumber \\
   & \quad \times \Bigg[ \frac{t_+ - t_-}{(m_W^2 - t_+)(m_W^2 - t_-)} \nonumber \\
   & \hspace{20mm} \times \left\{ 4 (s - m_{\eta^\pm}^2 - m_\ell^2) (s + m_W^2 - m_\ell^2 - m_{\eta^0}^2) + 4 m_{\eta^\pm}^2 (m_W^2 - m^2_\ell) \right. \nonumber \\
   & \hspace{26mm} \left. + 4 m_\ell^2  (1 + \delta_\eta) (s + m_W^2 - m_\ell^2 - m_{\eta^0}^2) - m_\ell^2 (1 + \delta_\eta)^2 (m_W^2 - m_\ell^2) \right\}  \nonumber \\
   & \qquad \quad + \ln \frac{m_W^2 - t_+}{m_W^2 - t_-} \cdot \left\{ 4 s + 4\delta_\eta m_\ell^2 - (1 + \delta_\eta)^2 m_\ell^2 \right\} \Bigg],
   \label{TAXS:dimensionfull}
\end{align}
where $t_\pm$ and $\delta_\eta$ are defined as
\begin{align}
   t_\pm &= 
   m_{\eta^\pm}^2 + m_{\eta^0}^2 - 2 E_{\eta^+} E_{\eta^0} \pm 2 |\vec{p}_{\eta^+}| |\vec{p}_{\eta^0}|~, \\
   \delta_\eta &=
   \frac{m_{\eta^\pm}^2 - m_{\eta^0}^2}{m_W^2} = \frac{\delta m_\eta (2 m_{\eta^\pm} - \delta m_\eta)}{m_W^2}~.
\end{align}

For $\eta_R t\rightarrow\sigma t$, the thermal averaged product of the scattering cross section and relative velocity $\langle\sigma v\rangle_{\eta_R t\rightarrow\sigma t}$ is
\begin{align}
\braket{\sigma v}_{\eta_R t\rightarrow \sigma t}&=\frac{1}{16 m^2_\eta m^2_t T K_2(m_\eta/T) K_2(m_t/T)} \nonumber \\
&\quad\times\int \frac{ds}{\sqrt{s}}K_1\left(\frac{\sqrt{s}}{T}\right)\sqrt{\{s-(m_t+m_\eta)^2\}\{s-(m_t-m_\eta)^2\}}[4\sigma v E_{t1}E_\eta] \nonumber \\
&=-\frac{\mu^2 y^2_t}{32\pi}\frac{z}{16m^3_\eta m^2_t K_2(z) K_2(r_t z)}\int \frac{ds}{\sqrt{s}}K_1\left(\frac{z}{m_\eta}\sqrt{s}\right) \nonumber \\
&\qquad\times\left[-\frac{\beta(4m^2_t-m^2_h)}{(\alpha+2m^2_t -m^2_h)^2-\beta^2}+\ln\frac{\alpha+\beta+2m^2_t -m^2_h}{\alpha-\beta+2m^2_t-m^2_h}\right],
\end{align}
where $\alpha$ and $\beta$ are defined as
\begin{align}
    \alpha&=-\frac{1}{2s}(s+m^2_t)(s+m^2_t-m^2_\eta), \\
    \beta&=\frac{1}{2s}(s-m^2_t)\sqrt{\{s-(m_t+m_\eta)^2\}\{s-(m_t-m_\eta)^2\}}.
\end{align}

\section{$\sigma$ production in freeze-in process}
\label{app:production}
In this section, we estimate the yield of $\sigma$ generated during the thermal-in regime.

\subsection{Analytic formula}
Since $\sigma$ is not thermalized at the initial ($z\lesssim 1$) but others are thermalized, each yield in the thermal-in regime can be approximated by $Y_\sigma\sim 0$ and $Y_0\sim Y_{0^*}\sim Y_{\pm} \sim Y_\eta^\textrm{EQ}$ in Eq.~(\ref{Boltzmann:ysigma}). As a result, we obtain the simplified evolution equation as
\begin{align}
    Y_\sigma'
    &\sim +\frac{2\gamma}{z}Y_\eta^\textrm{EQ}, \label{eq:approximated_eq_for_sigma}
\end{align}
where
\begin{align}
    \gamma
    &\equiv \left(1+\frac{1}{3}\frac{d\ln h(T)}{d\ln T}\right)\frac{\langle\Gamma_{\eta^0\to\sigma h}\rangle+\langle\Gamma_{\eta^0\to\sigma Z}\rangle+\langle\Gamma_{\eta^+\to\sigma W^+}\rangle+n_t^\textrm{EQ}\langle\sigma_{\eta^0 t\to\sigma t}\rangle}{H} \\
    &\sim \frac{4\langle\Gamma_{\eta^0\to\sigma h}\rangle+n_t^\textrm{EQ}\langle\sigma_{\eta^0 t\to\sigma t}\rangle}{H}.
\end{align}
Here we approximate $d \ln h(T)/d \ln T \sim 0$, $\langle\Gamma_{\eta^0\to\sigma h}\rangle\sim\langle\Gamma_{\eta^0\to\sigma Z}\rangle\sim\frac{1}{2}\langle\Gamma_{\eta^+\to\sigma W^+}\rangle$ and
\begin{align}
    Y_0\sim Y_{0^*}\sim Y_\pm \sim Y_\eta^\textrm{EQ} = \frac{m_\eta^3}{2\pi^2}\frac{1}{z}K_2(z)\cdot \frac{1}{s(T)}.
\end{align}
The solution of Eq.~(\ref{eq:approximated_eq_for_sigma}) is given by
\begin{align}
    Y_\sigma(z)
    &= \int_0^zdz'\frac{2\gamma(z')}{z'}Y_\eta^\textrm{EQ}(z') \\
    &\equiv Y_\sigma^{(\mathcal{D})}(z)+Y_\sigma^{(\mathcal{S})}(z), 
\end{align}
where we separated the contribution to $Y_\sigma$ into $Y_\sigma^{(\mathcal{D})}(z)$ and $Y_\sigma^{(\mathcal{S})}$, which are the contributions of the decay and the scattering processes, respectively.

At $z\rightarrow \infty$, the contribution of the decay can be estimated as
\begin{align}
    Y_\sigma^{(\mathcal{D})}(\infty)
    &\sim \int_0^\infty dz\frac{2\cdot 4\langle\Gamma_{\eta^0\to\sigma h}\rangle}{zH(T)}Y_\eta^\textrm{EQ}(z) \\
    &= \frac{3}{32\pi^2}\frac{\mu^2 m_\eta^2}{H(T=m_\eta)s(T=m_\eta)}. \label{eq:thermal-in_sigma_by_decay}
\end{align}
Note that the above result is proportional to $\mu^2/m_\eta^3$ since $H(T=m_\eta)\propto m^2_\eta$, and $s(T=m_\eta)\propto m^3_\eta$.
On the other hand, the contribution of the scattering can be estimated as
\begin{align}
    Y_\sigma^{(\mathcal{S})}(\infty)
    &\sim \int_0^\infty dz\frac{2n_t^\textrm{EQ}\langle\sigma_{\eta^0 t\to\sigma t}v\rangle}{zH(T)}Y_\eta^\textrm{EQ}(z) \\
    &\sim \frac{y_t^2}{16\pi^2}Y_\sigma^{(\mathcal{D})}(\infty)\cdot C_{\mathcal{S}},
\end{align}
where
\begin{align}
    C_{\mathcal{S}}(m_\eta, m_\sigma)
    \equiv \frac{16\pi}{\mu^2 m_\eta^2 y_t^2} \int_{(m_\eta+m_t)^2}^\infty ds \: \left(\frac{m_\eta}{\sqrt{s}}\right)^5\left(s-(m_\eta+m_t)^2\right)\left(s-(m_\eta-m_t)^2\right)\sigma_{\eta^0 t\to\sigma t}(s) \nonumber \\
\end{align}
is a dimensionless constant. In the case of $m_\eta = 1$ TeV $\gg m_\sigma$, one obtains
\begin{align}
    C_S(m_\eta=1 \ {\rm TeV})\sim0.43.
\end{align}
Therefore, we can conclude $Y_\sigma^{\rm thermal\mathchar`-in}\sim Y_\sigma(\infty)\sim Y_\sigma^{(\mathcal{D})}(\infty)\gg Y_\sigma^{(\mathcal{S})}(\infty)$.

\subsection{Constraint}
Combining the analytic result (\ref{eq:thermal-in_sigma_by_decay}) and the definition of the ratio $r=(Y_{\sigma,{\rm now}}-Y_{\Delta L})/Y_{\Delta L} \sim Y_\sigma^{(\mathcal{D})}(\infty)/Y_{\Delta L}$, we obtain a constraint on $m_\eta$-$\mu$ plane as
\begin{align}
\label{eq:result-approx}
    \frac{\mu^2}{m_\eta^3}
    \sim \frac{1.53\times 10^{-24}}{\rm GeV} r \times \sqrt{\frac{g(T=m_\eta)}{100}}\frac{h(T=m_\eta)}{100},
\end{align}
where $Y_{\Delta L}=2.44\times 10^{-10}$ is applied.
This constraint corresponds to the upward-sloping straight lines in Figure~\ref{fig:BBN:1MeV}.

{\small
\bibliographystyle{utphys28mod}
\bibliography{ref}

\providecommand{\href}[2]{#2}\begingroup\begin{thebibliography}{10}

\bibitem{Planck:2018vyg}
{\bfseries Planck} Collaboration, ``{Planck 2018 results. VI. Cosmological
  parameters},'' \href{https://dx.doi.org/10.1051/0004-6361/201833910}{Astron.\
   Astrophys.\  {\bfseries 641} (2020) A6} {\ttfamily
  [\href{https://arxiv.org/abs/1807.06209}{arXiv:1807.06209}]}. [Erratum:
  Astron.Astrophys. 652, C4 (2021)].

\bibitem{WMAP:2012nax}
{\bfseries WMAP} Collaboration, ``{Nine-Year Wilkinson Microwave Anisotropy
  Probe (WMAP) Observations: Cosmological Parameter Results},''
  \href{https://dx.doi.org/10.1088/0067-0049/208/2/19}{Astrophys.\  J.\
  Suppl.\  {\bfseries 208} (2013) 19} {\ttfamily
  [\href{https://arxiv.org/abs/1212.5226}{arXiv:1212.5226}]}.

\bibitem{Minkowski:1977sc}
P.~Minkowski, ``{$\mu \to e\gamma$ at a Rate of One Out of $10^{9}$ Muon
  Decays?}'' \href{https://dx.doi.org/10.1016/0370-2693(77)90435-X}{Phys.\
  Lett.\  B {\bfseries 67} (1977) 421--428}.

\bibitem{Yanagida:1979as}
T.~Yanagida, ``{HORIZONTAL SYMMETRY AND MASSES OF NEUTRINOS},''
Conf.\  Proc.\  {\bfseries C7902131} (1979) 95--99.

\bibitem{Gell-Mann:1979vob}
M.~Gell-Mann, P.~Ramond, and R.~Slansky, ``{Complex Spinors and Unified
  Theories},'' Conf.\  Proc.\  C {\bfseries 790927} (1979) 315--321 {\ttfamily
  [\href{https://arxiv.org/abs/1306.4669}{arXiv:1306.4669}]}.

\bibitem{Mohapatra:1979ia}
R.~N.~Mohapatra and G.~Senjanovic, ``{Neutrino Mass and Spontaneous Parity
  Nonconservation},''
  \href{https://dx.doi.org/10.1103/PhysRevLett.44.912}{Phys.\  Rev.\  Lett.\
  {\bfseries 44} (1980) 912}.

\bibitem{Fukugita:1986hr}
M.~Fukugita and T.~Yanagida, ``{Baryogenesis Without Grand Unification},''
  \href{https://dx.doi.org/10.1016/0370-2693(86)91126-3}{Phys.\  Lett.\  B
  {\bfseries 174} (1986) 45--47}.

\bibitem{Kuzmin:1985mm}
V.~A.~Kuzmin, V.~A.~Rubakov, and M.~E.~Shaposhnikov, ``{On the Anomalous
  Electroweak Baryon Number Nonconservation in the Early Universe},''
  \href{https://dx.doi.org/10.1016/0370-2693(85)91028-7}{Phys.\  Lett.\  B
  {\bfseries 155} (1985) 36}.

\bibitem{Tao:1996vb}
Z.-j.~Tao, ``{Radiative seesaw mechanism at weak scale},''
  \href{https://dx.doi.org/10.1103/PhysRevD.54.5693}{Phys.\  Rev.\  D
  {\bfseries 54} (1996) 5693--5697} {\ttfamily
  [\href{https://arxiv.org/abs/hep-ph/9603309}{hep-ph/9603309}]}.

\bibitem{Ma:2006km}
E.~Ma, ``{Verifiable radiative seesaw mechanism of neutrino mass and dark
  matter},'' \href{https://dx.doi.org/10.1103/PhysRevD.73.077301}{Phys.\  Rev.\
   D {\bfseries 73} (2006) 077301} {\ttfamily
  [\href{https://arxiv.org/abs/hep-ph/0601225}{hep-ph/0601225}]}.

\bibitem{Hugle:2018qbw}
T.~Hugle, M.~Platscher, and K.~Schmitz, ``{Low-Scale Leptogenesis in the
  Scotogenic Neutrino Mass Model},''
  \href{https://dx.doi.org/10.1103/PhysRevD.98.023020}{Phys.\  Rev.\  D
  {\bfseries 98} (2018) 023020} {\ttfamily
  [\href{https://arxiv.org/abs/1804.09660}{arXiv:1804.09660}]}.

\bibitem{Borah:2018uci}
D.~Borah, A.~Dasgupta, and S.~K.~Kang, ``{TeV Scale Leptogenesis via Dark
  Sector Scatterings},''
  \href{https://dx.doi.org/10.1140/epjc/s10052-020-8052-1}{Eur.\  Phys.\  J.\
  C {\bfseries 80} (2020) 498} {\ttfamily
  [\href{https://arxiv.org/abs/1806.04689}{arXiv:1806.04689}]}.

\bibitem{Bose:2024bnp}
D.~Bose, R.~Pramanick, and T.~S.~Ray, ``{Cogenesis of visible and dark matter
  in a scotogenic model}.'' {\ttfamily
  \href{https://arxiv.org/abs/2409.06541}{arXiv:2409.06541}}.

\bibitem{Racker:2024fpn}
J.~Racker, ``{Low-scale leptogenesis in the scotogenic model: Spectator
  processes and benchmark points},''
  \href{https://dx.doi.org/10.1103/PhysRevD.111.L081301}{Phys.\  Rev.\  D
  {\bfseries 111} (2025) L081301} {\ttfamily
  [\href{https://arxiv.org/abs/2411.15120}{arXiv:2411.15120}]}.

\bibitem{Nussinov:1985xr}
S.~Nussinov, ``{TECHNOCOSMOLOGY: COULD A TECHNIBARYON EXCESS PROVIDE A
  'NATURAL' MISSING MASS CANDIDATE?}''
  \href{https://dx.doi.org/10.1016/0370-2693(85)90689-6}{Phys.\  Lett.\  B
  {\bfseries 165} (1985) 55--58}.

\bibitem{Barr:1990ca}
S.~M.~Barr, R.~S.~Chivukula, and E.~Farhi, ``{Electroweak Fermion Number
  Violation and the Production of Stable Particles in the Early Universe},''
  \href{https://dx.doi.org/10.1016/0370-2693(90)91661-T}{Phys.\  Lett.\  B
  {\bfseries 241} (1990) 387--391}.

\bibitem{Barr:1991qn}
S.~M.~Barr, ``{Baryogenesis, sphalerons and the cogeneration of dark matter},''
  \href{https://dx.doi.org/10.1103/PhysRevD.44.3062}{Phys.\  Rev.\  D
  {\bfseries 44} (1991) 3062--3066}.

\bibitem{Dodelson:1991iv}
S.~Dodelson, B.~R.~Greene, and L.~M.~Widrow, ``{Baryogenesis, dark matter and
  the width of the Z},''
  \href{https://dx.doi.org/10.1016/0550-3213(92)90328-9}{Nucl.\  Phys.\  B
  {\bfseries 372} (1992) 467--493}.

\bibitem{Kaplan:1991ah}
D.~B.~Kaplan, ``{A Single explanation for both the baryon and dark matter
  densities},'' \href{https://dx.doi.org/10.1103/PhysRevLett.68.741}{Phys.\
  Rev.\  Lett.\  {\bfseries 68} (1992) 741--743}.

\bibitem{Kuzmin:1996he}
S.~J.~Ball and Y.~A.~Kamyshkov, eds., ``{A Simultaneous solution to
  baryogenesis and dark matter problems},''
  \href{https://dx.doi.org/10.1134/1.953070}{Phys.\  Part.\  Nucl.\  {\bfseries
  29} (1998) 257--265} {\ttfamily
  [\href{https://arxiv.org/abs/hep-ph/9701269}{hep-ph/9701269}]}.

\bibitem{Foot:2003jt}
R.~Foot and R.~R.~Volkas, ``{Was ordinary matter synthesized from mirror
  matter? An Attempt to explain why Omega(Baryon) approximately equal to 0.2
  Omega(Dark)},'' \href{https://dx.doi.org/10.1103/PhysRevD.68.021304}{Phys.\
  Rev.\  D {\bfseries 68} (2003) 021304} {\ttfamily
  [\href{https://arxiv.org/abs/hep-ph/0304261}{hep-ph/0304261}]}.

\bibitem{Foot:2004pq}
R.~Foot and R.~R.~Volkas, ``{Explaining Omega(Baryon) approximately 0.2
  Omega(Dark) through the synthesis of ordinary matter from mirror matter: A
  More general analysis},''
  \href{https://dx.doi.org/10.1103/PhysRevD.69.123510}{Phys.\  Rev.\  D
  {\bfseries 69} (2004) 123510} {\ttfamily
  [\href{https://arxiv.org/abs/hep-ph/0402267}{hep-ph/0402267}]}.

\bibitem{Hooper:2004dc}
D.~Hooper, J.~March-Russell, and S.~M.~West, ``{Asymmetric sneutrino dark
  matter and the Omega(b) / Omega(DM) puzzle},''
  \href{https://dx.doi.org/10.1016/j.physletb.2004.11.047}{Phys.\  Lett.\  B
  {\bfseries 605} (2005) 228--236} {\ttfamily
  [\href{https://arxiv.org/abs/hep-ph/0410114}{hep-ph/0410114}]}.

\bibitem{Kitano:2004sv}
R.~Kitano and I.~Low, ``{Dark matter from baryon asymmetry},''
  \href{https://dx.doi.org/10.1103/PhysRevD.71.023510}{Phys.\  Rev.\  D
  {\bfseries 71} (2005) 023510} {\ttfamily
  [\href{https://arxiv.org/abs/hep-ph/0411133}{hep-ph/0411133}]}.

\bibitem{Gudnason:2006ug}
S.~B.~Gudnason, C.~Kouvaris, and F.~Sannino, ``{Towards working technicolor:
  Effective theories and dark matter},''
  \href{https://dx.doi.org/10.1103/PhysRevD.73.115003}{Phys.\  Rev.\  D
  {\bfseries 73} (2006) 115003} {\ttfamily
  [\href{https://arxiv.org/abs/hep-ph/0603014}{hep-ph/0603014}]}.

\bibitem{Kaplan:2009ag}
D.~E.~Kaplan, M.~A.~Luty, and K.~M.~Zurek, ``{Asymmetric Dark Matter},''
  \href{https://dx.doi.org/10.1103/PhysRevD.79.115016}{Phys.\  Rev.\  D
  {\bfseries 79} (2009) 115016} {\ttfamily
  [\href{https://arxiv.org/abs/0901.4117}{arXiv:0901.4117}]}.

\bibitem{Davoudiasl:2012uw}
H.~Davoudiasl and R.~N.~Mohapatra, ``{On Relating the Genesis of Cosmic Baryons
  and Dark Matter},''
  \href{https://dx.doi.org/10.1088/1367-2630/14/9/095011}{New J.\  Phys.\
  {\bfseries 14} (2012) 095011} {\ttfamily
  [\href{https://arxiv.org/abs/1203.1247}{arXiv:1203.1247}]}.

\bibitem{Petraki:2013wwa}
K.~Petraki and R.~R.~Volkas, ``{Review of asymmetric dark matter},''
  \href{https://dx.doi.org/10.1142/S0217751X13300287}{Int.\  J.\  Mod.\  Phys.\
   A {\bfseries 28} (2013) 1330028} {\ttfamily
  [\href{https://arxiv.org/abs/1305.4939}{arXiv:1305.4939}]}.

\bibitem{Zurek:2013wia}
K.~M.~Zurek, ``{Asymmetric Dark Matter: Theories, Signatures, and
  Constraints},''
  \href{https://dx.doi.org/10.1016/j.physrep.2013.12.001}{Phys.\  Rept.\
  {\bfseries 537} (2014) 91--121} {\ttfamily
  [\href{https://arxiv.org/abs/1308.0338}{arXiv:1308.0338}]}.

\bibitem{Asai:2022vat}
K.~Asai, Y.~Sakai, J.~Sato, Y.~Takanishi, and M.~Yamanaka, ``{Asymmetric
  mediator in scotogenic model},''
  \href{https://dx.doi.org/10.1016/j.physletb.2022.137627}{Phys.\  Lett.\  B
  {\bfseries 836} (2023) 137627} {\ttfamily
  [\href{https://arxiv.org/abs/2209.08257}{arXiv:2209.08257}]}.

\bibitem{Reno:1987qw}
M.~H.~Reno and D.~Seckel, ``{Primordial Nucleosynthesis: The Effects of
  Injecting Hadrons},''
  \href{https://dx.doi.org/10.1103/PhysRevD.37.3441}{Phys.\  Rev.\  D
  {\bfseries 37} (1988) 3441}.

\bibitem{Kawasaki:2004qu}
M.~Kawasaki, K.~Kohri, and T.~Moroi, ``{Big-Bang nucleosynthesis and hadronic
  decay of long-lived massive particles},''
  \href{https://dx.doi.org/10.1103/PhysRevD.71.083502}{Phys.\  Rev.\  D
  {\bfseries 71} (2005) 083502} {\ttfamily
  [\href{https://arxiv.org/abs/astro-ph/0408426}{astro-ph/0408426}]}.

\bibitem{Kawasaki:2020qxm}
M.~Kawasaki, K.~Kohri, T.~Moroi, K.~Murai, and H.~Murayama, ``{Big-bang
  nucleosynthesis with sub-GeV massive decaying particles},''
  \href{https://dx.doi.org/10.1088/1475-7516/2020/12/048}{JCAP {\bfseries 12}
  (2020) 048} {\ttfamily
  [\href{https://arxiv.org/abs/2006.14803}{arXiv:2006.14803}]}.

\bibitem{Kohri:2001jx}
K.~Kohri, ``{Primordial nucleosynthesis and hadronic decay of a massive
  particle with a relatively short lifetime},''
  \href{https://dx.doi.org/10.1103/PhysRevD.64.043515}{Phys.\  Rev.\  D
  {\bfseries 64} (2001) 043515} {\ttfamily
  [\href{https://arxiv.org/abs/astro-ph/0103411}{astro-ph/0103411}]}.

\bibitem{Kawasaki:2017bqm}
M.~Kawasaki, K.~Kohri, T.~Moroi, and Y.~Takaesu, ``{Revisiting Big-Bang
  Nucleosynthesis Constraints on Long-Lived Decaying Particles},''
  \href{https://dx.doi.org/10.1103/PhysRevD.97.023502}{Phys.\  Rev.\  D
  {\bfseries 97} (2018) 023502} {\ttfamily
  [\href{https://arxiv.org/abs/1709.01211}{arXiv:1709.01211}]}.

\bibitem{Jedamzik:2006xz}
K.~Jedamzik, ``{Big bang nucleosynthesis constraints on hadronically and
  electromagnetically decaying relic neutral particles},''
  \href{https://dx.doi.org/10.1103/PhysRevD.74.103509}{Phys.\  Rev.\  D
  {\bfseries 74} (2006) 103509} {\ttfamily
  [\href{https://arxiv.org/abs/hep-ph/0604251}{hep-ph/0604251}]}.

\bibitem{Khlopov:1984pf}
M.~Y.~Khlopov and A.~D.~Linde, ``{Is It Easy to Save the Gravitino?}''
  \href{https://dx.doi.org/10.1016/0370-2693(84)91656-3}{Phys.\  Lett.\  B
  {\bfseries 138} (1984) 265--268}.

\bibitem{Lindley:1984bg}
D.~Lindley, ``{Cosmological Constraints on the Lifetime of Massive
  Particles},'' \href{https://dx.doi.org/10.1086/163267}{Astrophys.\  J.\
  {\bfseries 294} (1985) 1--8}.

\bibitem{Forestell:2018txr}
L.~Forestell, D.~E.~Morrissey, and G.~White, ``{Limits from BBN on Light
  Electromagnetic Decays},''
  \href{https://dx.doi.org/10.1007/JHEP01(2019)074}{JHEP {\bfseries 01} (2019)
  074} {\ttfamily [\href{https://arxiv.org/abs/1809.01179}{arXiv:1809.01179}]}.

\bibitem{Dimopoulos:1988ue}
S.~Dimopoulos, R.~Esmailzadeh, L.~J.~Hall, and G.~D.~Starkman, ``{Limits on
  Late Decaying Particles From Nucleosynthesis},''
  \href{https://dx.doi.org/10.1016/0550-3213(89)90173-9}{Nucl.\  Phys.\  B
  {\bfseries 311} (1989) 699--718}.

\bibitem{Jedamzik:2004er}
K.~Jedamzik, ``{Did something decay, evaporate, or annihilate during Big Bang
  nucleosynthesis?}''
  \href{https://dx.doi.org/10.1103/PhysRevD.70.063524}{Phys.\  Rev.\  D
  {\bfseries 70} (2004) 063524} {\ttfamily
  [\href{https://arxiv.org/abs/astro-ph/0402344}{astro-ph/0402344}]}.

\bibitem{Cumberbatch:2007me}
D.~Cumberbatch, K.~Ichikawa, M.~Kawasaki, K.~Kohri, \emph{et al}., ``{Solving
  the cosmic lithium problems with primordial late-decaying particles},''
  \href{https://dx.doi.org/10.1103/PhysRevD.76.123005}{Phys.\  Rev.\  D
  {\bfseries 76} (2007) 123005} {\ttfamily
  [\href{https://arxiv.org/abs/0708.0095}{arXiv:0708.0095}]}.

\bibitem{Koren:2022axd}
S.~Koren, ``{Cosmological Lithium Solution from Discrete Gauged B-L},''
  \href{https://dx.doi.org/10.1103/PhysRevLett.131.091003}{Phys.\  Rev.\
  Lett.\  {\bfseries 131} (2023) 091003} {\ttfamily
  [\href{https://arxiv.org/abs/2204.01750}{arXiv:2204.01750}]}.

\bibitem{Goudelis:2015wpa}
A.~Goudelis, M.~Pospelov, and J.~Pradler, ``{Light Particle Solution to the
  Cosmic Lithium Problem},''
  \href{https://dx.doi.org/10.1103/PhysRevLett.116.211303}{Phys.\  Rev.\
  Lett.\  {\bfseries 116} (2016) 211303} {\ttfamily
  [\href{https://arxiv.org/abs/1510.08858}{arXiv:1510.08858}]}.

\bibitem{Yamazaki:2014fja}
D.~G.~Yamazaki, M.~Kusakabe, T.~Kajino, G.~J.~Mathews, and M.-K.~Cheoun,
  ``{Cosmological solutions to the Lithium problem: Big-bang nucleosynthesis
  with photon cooling, $X$-particle decay and a primordial magnetic field},''
  \href{https://dx.doi.org/10.1103/PhysRevD.90.023001}{Phys.\  Rev.\  D
  {\bfseries 90} (2014) 023001} {\ttfamily
  [\href{https://arxiv.org/abs/1407.0021}{arXiv:1407.0021}]}.

\bibitem{ParticleDataGroup:2024cfk}
{\bfseries Particle Data Group} Collaboration, ``{Review of particle
  physics},'' \href{https://dx.doi.org/10.1103/PhysRevD.110.030001}{Phys.\
  Rev.\  D {\bfseries 110} (2024) 030001}.

\bibitem{Pospelov:2006sc}
M.~Pospelov, ``{Particle physics catalysis of thermal Big Bang
  Nucleosynthesis},''
  \href{https://dx.doi.org/10.1103/PhysRevLett.98.231301}{Phys.\  Rev.\  Lett.\
   {\bfseries 98} (2007) 231301} {\ttfamily
  [\href{https://arxiv.org/abs/hep-ph/0605215}{hep-ph/0605215}]}.

\bibitem{Pospelov:2008ta}
M.~Pospelov, J.~Pradler, and F.~D.~Steffen, ``{Constraints on Supersymmetric
  Models from Catalytic Primordial Nucleosynthesis of Beryllium},''
  \href{https://dx.doi.org/10.1088/1475-7516/2008/11/020}{JCAP {\bfseries 11}
  (2008) 020} {\ttfamily
  [\href{https://arxiv.org/abs/0807.4287}{arXiv:0807.4287}]}.

\bibitem{Khlopov:2007ic}
M.~Y.~Khlopov and C.~Kouvaris, ``{Strong Interactive Massive Particles from a
  Strong Coupled Theory},''
  \href{https://dx.doi.org/10.1103/PhysRevD.77.065002}{Phys.\  Rev.\  D
  {\bfseries 77} (2008) 065002} {\ttfamily
  [\href{https://arxiv.org/abs/0710.2189}{arXiv:0710.2189}]}.

\bibitem{Jittoh:2011ni}
T.~Jittoh, K.~Kohri, M.~Koike, J.~Sato, \emph{et al}., ``{Big-bang
  nucleosynthesis with a long-lived charged massive particle including $^4$He
  spallation processes},''
  \href{https://dx.doi.org/10.1103/PhysRevD.84.035008}{Phys.\  Rev.\  D
  {\bfseries 84} (2011) 035008} {\ttfamily
  [\href{https://arxiv.org/abs/1105.1431}{arXiv:1105.1431}]}.

\bibitem{Kamimura:2008fx}
M.~Kamimura, Y.~Kino, and E.~Hiyama, ``{Big-Bang Nucleosynthesis Reactions
  Catalyzed by a Long-Lived Negatively-Charged Leptonic Particle},''
  \href{https://dx.doi.org/10.1143/PTP.121.1059}{Prog.\  Theor.\  Phys.\
  {\bfseries 121} (2009) 1059--1098} {\ttfamily
  [\href{https://arxiv.org/abs/0809.4772}{arXiv:0809.4772}]}.

\bibitem{Kaplinghat:2006qr}
M.~Kaplinghat and A.~Rajaraman, ``{Big Bang Nucleosynthesis with Bound States
  of Long-lived Charged Particles},''
  \href{https://dx.doi.org/10.1103/PhysRevD.74.103004}{Phys.\  Rev.\  D
  {\bfseries 74} (2006) 103004} {\ttfamily
  [\href{https://arxiv.org/abs/astro-ph/0606209}{astro-ph/0606209}]}.

\bibitem{Kohri:2006cn}
K.~Kohri and F.~Takayama, ``{Big bang nucleosynthesis with long lived charged
  massive particles},''
  \href{https://dx.doi.org/10.1103/PhysRevD.76.063507}{Phys.\  Rev.\  D
  {\bfseries 76} (2007) 063507} {\ttfamily
  [\href{https://arxiv.org/abs/hep-ph/0605243}{hep-ph/0605243}]}.

\bibitem{Hamaguchi:2007mp}
K.~Hamaguchi, T.~Hatsuda, M.~Kamimura, Y.~Kino, and T.~T.~Yanagida,
  ``{Stau-catalyzed $^{6}Li$ Production in Big-Bang Nucleosynthesis},''
  \href{https://dx.doi.org/10.1016/j.physletb.2007.05.030}{Phys.\  Lett.\  B
  {\bfseries 650} (2007) 268--274} {\ttfamily
  [\href{https://arxiv.org/abs/hep-ph/0702274}{hep-ph/0702274}]}.

\bibitem{Jittoh:2007fr}
T.~Jittoh, K.~Kohri, M.~Koike, J.~Sato, \emph{et al}., ``{Possible solution to
  the Li-7 problem by the long lived stau},''
  \href{https://dx.doi.org/10.1103/PhysRevD.76.125023}{Phys.\  Rev.\  D
  {\bfseries 76} (2007) 125023} {\ttfamily
  [\href{https://arxiv.org/abs/0704.2914}{arXiv:0704.2914}]}.

\bibitem{Jittoh:2010wh}
T.~Jittoh, K.~Kohri, M.~Koike, J.~Sato, \emph{et al}., ``{Stau relic density at
  the Big-Bang nucleosynthesis era consistent with the abundance of the light
  element nuclei in the coannihilation scenario},''
  \href{https://dx.doi.org/10.1103/PhysRevD.82.115030}{Phys.\  Rev.\  D
  {\bfseries 82} (2010) 115030} {\ttfamily
  [\href{https://arxiv.org/abs/1001.1217}{arXiv:1001.1217}]}.

\bibitem{Jittoh:2008eq}
T.~Jittoh, K.~Kohri, M.~Koike, J.~Sato, \emph{et al}., ``{Big-bang
  nucleosynthesis and the relic abundance of dark matter in a stau-neutralino
  coannihilation scenario},''
  \href{https://dx.doi.org/10.1103/PhysRevD.78.055007}{Phys.\  Rev.\  D
  {\bfseries 78} (2008) 055007} {\ttfamily
  [\href{https://arxiv.org/abs/0805.3389}{arXiv:0805.3389}]}.

\bibitem{Bird:2007ge}
C.~Bird, K.~Koopmans, and M.~Pospelov, ``{Primordial Lithium Abundance in
  Catalyzed Big Bang Nucleosynthesis},''
  \href{https://dx.doi.org/10.1103/PhysRevD.78.083010}{Phys.\  Rev.\  D
  {\bfseries 78} (2008) 083010} {\ttfamily
  [\href{https://arxiv.org/abs/hep-ph/0703096}{hep-ph/0703096}]}.

\bibitem{Jedamzik:2007cp}
K.~Jedamzik, ``{The cosmic Li-6 and Li-7 problems and BBN with long-lived
  charged massive particles},''
  \href{https://dx.doi.org/10.1103/PhysRevD.77.063524}{Phys.\  Rev.\  D
  {\bfseries 77} (2008) 063524} {\ttfamily
  [\href{https://arxiv.org/abs/0707.2070}{arXiv:0707.2070}]}.

\bibitem{Kusakabe:2007fu}
M.~Kusakabe, T.~Kajino, R.~N.~Boyd, T.~Yoshida, and G.~J.~Mathews, ``{A
  Simultaneous Solution to the {\textasciicircum}6Li and {\textasciicircum}7Li
  Big Bang Nucleosynthesis Problems from a Long-Lived Negatively-Charged
  Leptonic Particle},''
  \href{https://dx.doi.org/10.1103/PhysRevD.76.121302}{Phys.\  Rev.\  D
  {\bfseries 76} (2007) 121302} {\ttfamily
  [\href{https://arxiv.org/abs/0711.3854}{arXiv:0711.3854}]}.

\bibitem{Bailly:2008yy}
S.~Bailly, K.~Jedamzik, and G.~Moultaka, ``{Gravitino Dark Matter and the
  Cosmic Lithium Abundances},''
  \href{https://dx.doi.org/10.1103/PhysRevD.80.063509}{Phys.\  Rev.\  D
  {\bfseries 80} (2009) 063509} {\ttfamily
  [\href{https://arxiv.org/abs/0812.0788}{arXiv:0812.0788}]}.

\bibitem{Boyd:2010kj}
R.~N.~Boyd, C.~R.~Brune, G.~M.~Fuller, and C.~J.~Smith, ``{New Nuclear Physics
  for Big Bang Nucleosynthesis},''
  \href{https://dx.doi.org/10.1103/PhysRevD.82.105005}{Phys.\  Rev.\  D
  {\bfseries 82} (2010) 105005} {\ttfamily
  [\href{https://arxiv.org/abs/1008.0848}{arXiv:1008.0848}]}.

\bibitem{LopezHonorez:2006gr}
L.~Lopez~Honorez, E.~Nezri, J.~F.~Oliver, and M.~H.~G.~Tytgat, ``{The Inert
  Doublet Model: An Archetype for Dark Matter},''
  \href{https://dx.doi.org/10.1088/1475-7516/2007/02/028}{JCAP {\bfseries 02}
  (2007) 028} {\ttfamily
  [\href{https://arxiv.org/abs/hep-ph/0612275}{hep-ph/0612275}]}.

\bibitem{Sagunski:2020spe}
L.~Sagunski, S.~Gad-Nasr, B.~Colquhoun, A.~Robertson, and S.~Tulin,
  ``{Velocity-dependent Self-interacting Dark Matter from Groups and Clusters
  of Galaxies},'' \href{https://dx.doi.org/10.1088/1475-7516/2021/01/024}{JCAP
  {\bfseries 01} (2021) 024} {\ttfamily
  [\href{https://arxiv.org/abs/2006.12515}{arXiv:2006.12515}]}.

\bibitem{Andrade:2020lqq}
K.~E.~Andrade, J.~Fuson, S.~Gad-Nasr, D.~Kong, \emph{et al}., ``{A stringent
  upper limit on dark matter self-interaction cross-section from cluster strong
  lensing},'' \href{https://dx.doi.org/10.1093/mnras/stab3241}{Mon.\  Not.\
  Roy.\  Astron.\  Soc.\  {\bfseries 510} (2021) 54--81} {\ttfamily
  [\href{https://arxiv.org/abs/2012.06611}{arXiv:2012.06611}]}.

\bibitem{Eckert:2022qia}
D.~Eckert, S.~Ettori, A.~Robertson, R.~Massey, \emph{et al}., ``{Constraints on
  dark matter self-interaction from the internal density profiles of X-COP
  galaxy clusters},''
  \href{https://dx.doi.org/10.1051/0004-6361/202243205}{Astron.\  Astrophys.\
  {\bfseries 666} (2022) A41} {\ttfamily
  [\href{https://arxiv.org/abs/2205.01123}{arXiv:2205.01123}]}.

\bibitem{K:2023huw}
G.~K. and S.~Desai, ``{Constraints on Self-Interacting dark matter from relaxed
  galaxy groups},'' \href{https://dx.doi.org/10.1016/j.dark.2023.101291}{Phys.\
   Dark Univ.\  {\bfseries 42} (2023) 101291} {\ttfamily
  [\href{https://arxiv.org/abs/2307.05880}{arXiv:2307.05880}]}.

\bibitem{ATLAS:2022pib}
{\bfseries ATLAS} Collaboration, ``{Search for heavy, long-lived, charged
  particles with large ionisation energy loss in $pp$ collisions at $\sqrt{s} =
  13~\text{TeV}$ using the ATLAS experiment and the full Run 2 dataset},''
  \href{https://dx.doi.org/10.1007/JHEP06(2023)158}{JHEP {\bfseries 2306}
  (2023) 158} {\ttfamily
  [\href{https://arxiv.org/abs/2205.06013}{arXiv:2205.06013}]}.

\bibitem{CMS:2024nhn}
{\bfseries CMS} Collaboration, ``{Search for heavy long-lived charged particles
  with large ionization energy loss in proton-proton collisions at $ \sqrt{s} $
  = 13 TeV},'' \href{https://dx.doi.org/10.1007/JHEP04(2025)109}{JHEP
  {\bfseries 04} (2025) 109} {\ttfamily
  [\href{https://arxiv.org/abs/2410.09164}{arXiv:2410.09164}]}.

\bibitem{ATLAS:2025fdm}
{\bfseries ATLAS} Collaboration, ``{Search for long-lived charged particles
  using large specific ionisation loss and time of flight in 140 fb$^{-1}$ of
  pp collisions at $\sqrt{s}$ = 13 TeV with the ATLAS detector},''
  \href{https://dx.doi.org/10.1007/JHEP07(2025)140}{JHEP {\bfseries 07} (2025)
  140} {\ttfamily [\href{https://arxiv.org/abs/2502.06694}{arXiv:2502.06694}]}.

\bibitem{Harvey:1990qw}
J.~A.~Harvey and M.~S.~Turner, ``{Cosmological baryon and lepton number in the
  presence of electroweak fermion number violation},''
  \href{https://dx.doi.org/10.1103/PhysRevD.42.3344}{Phys.\  Rev.\  D
  {\bfseries 42} (1990) 3344--3349}.

\bibitem{Kolb:1990vq}
E.~W.~Kolb and M.~S.~Turner,
  \href{https://dx.doi.org/10.1201/9780429492860}{{\em {The Early Universe}}},
  vol.~69.
\newblock Taylor and Francis, 2019.

\bibitem{Enomoto:2023cun}
S.~Enomoto, Y.-H.~Su, M.-Z.~Zheng, and H.-H.~Zhang, ``{Boltzmann Equation and
  Its Cosmological Applications},''
  \href{https://dx.doi.org/10.3390/sym17060921}{Symmetry {\bfseries 17} (2025)
  921} {\ttfamily [\href{https://arxiv.org/abs/2301.11819}{arXiv:2301.11819}]}.

\bibitem{Avila:2021mwg}
I.~M.~\'Avila, G.~Cottin, and M.~A.~D\'\i{}az, ``{Revisiting the scotogenic
  model with scalar dark matter},''
  \href{https://dx.doi.org/10.1088/1361-6471/ac5fb4}{J.\  Phys.\  G {\bfseries
  49} (2022) 065001} {\ttfamily
  [\href{https://arxiv.org/abs/2108.05103}{arXiv:2108.05103}]}.

\bibitem{Gondolo:1990dk}
P.~Gondolo and G.~Gelmini, ``{Cosmic abundances of stable particles: Improved
  analysis},'' \href{https://dx.doi.org/10.1016/0550-3213(91)90438-4}{Nucl.\
  Phys.\  B {\bfseries 360} (1991) 145--179}.

\bibitem{Drees:2015exa}
M.~Drees, F.~Hajkarim, and E.~R.~Schmitz, ``{The Effects of QCD Equation of
  State on the Relic Density of WIMP Dark Matter},''
  \href{https://dx.doi.org/10.1088/1475-7516/2015/06/025}{JCAP {\bfseries 06}
  (2015) 025} {\ttfamily
  [\href{https://arxiv.org/abs/1503.03513}{arXiv:1503.03513}]}.

\bibitem{Saikawa:2020swg}
K.~Saikawa and S.~Shirai, ``{Precise WIMP Dark Matter Abundance and Standard
  Model Thermodynamics},''
  \href{https://dx.doi.org/10.1088/1475-7516/2020/08/011}{JCAP {\bfseries 08}
  (2020) 011} {\ttfamily
  [\href{https://arxiv.org/abs/2005.03544}{arXiv:2005.03544}]}.

\bibitem{Fields:1998gv}
B.~D.~Fields and K.~A.~Olive, ``{On the evolution of helium in blue compact
  galaxies},'' \href{https://dx.doi.org/10.1086/306248}{Astrophys.\  J.\
  {\bfseries 506} (1998) 177} {\ttfamily
  [\href{https://arxiv.org/abs/astro-ph/9803297}{astro-ph/9803297}]}.

\bibitem{Izotov:2003xn}
Y.~I.~Izotov and T.~X.~Thuan, ``{Systematic effects and a new determination of
  the primordial abundance of He-4 and dY/dZ from observations of blue compact
  galaxies},'' \href{https://dx.doi.org/10.1086/380830}{Astrophys.\  J.\
  {\bfseries 602} (2004) 200--230} {\ttfamily
  [\href{https://arxiv.org/abs/astro-ph/0310421}{astro-ph/0310421}]}.

\bibitem{Matsumoto:2022tlr}
A.~Matsumoto \emph{et al}., ``{EMPRESS. VIII. A New Determination of Primordial
  He Abundance with Extremely Metal-poor Galaxies: A Suggestion of the Lepton
  Asymmetry and Implications for the Hubble Tension},''
  \href{https://dx.doi.org/10.3847/1538-4357/ac9ea1}{Astrophys.\  J.\
  {\bfseries 941} (2022) 167} {\ttfamily
  [\href{https://arxiv.org/abs/2203.09617}{arXiv:2203.09617}]}.

\end{thebibliography}\endgroup
}

\end{document}